\documentclass[useAMS,usenatbib]{mn2e}
\usepackage{graphicx}
\usepackage{amssymb}
\usepackage{amsmath}
\usepackage[utf8]{inputenc}
\usepackage{color}
\usepackage{multirow}
\newcommand{\apj}{ApJ}
\newcommand{\apjl}{ApJL}
\newcommand{\apss}{Ap\&SS}

\newcommand{\araa}{ARA\&A}
\newcommand{\aap}{A\&A}
\newcommand{\aj}{AJ}
\newcommand{\mnras}{MNRAS}
\newcommand{\nat}{Nature}

\newcommand{\asprv}{ASPRv}
\newcommand{\pasj}{PASJ}
\newcommand{\ssr}{Space Sci. Rev.}

\def\XMM{\mbox{\textit{XMM-Newton}}}
\def\SS433{\mbox{SS\,433}}
\def\6946{\mbox{NGC\,6946~X-1}}
\def\5408{\mbox{NGC\,5408~X-1}}
\def\M82{\mbox{M\,82~X-1}}
\def\IC342{\mbox{IC\,342~X-1}}

\def\1313{\mbox{NGC\,1313~X-1}}
\begin{document}
\title[ULXs with flat-topped noise and QPO]
{Ultraluminous X-ray sources with flat-topped noise and QPO}

\author[K. Atapin, S. Fabrika \& M. D. Caballero-Garc\'{\i}a]
{Kirill Atapin,$^{1,2}$\thanks{E-mail: atapin.kirill@gmail.com} 
Sergei Fabrika,$^{2,3}$
Maria~D. Caballero-Garc\'{\i}a$^{4}$\\
$^1$ Sternberg Astronomical Institute, Moscow State University, Universitetsky pr., 13, Moscow, 119991, Russia\\
$^2$ Special Astrophysical Observatory, Nizhnij Arkhyz, 369167, Russia\\
$^3$ Kazan Federal University, Kremlevskaya 18, Kazan 420008, Russia\\
$^4$ Astronomical Institute, Academy of Sciences, Bo\v{c}n\'{\i}~II~1401, CZ-14100~Prague, Czech~Republic}

\pagerange{\pageref{firstpage}--\pageref{lastpage}}
\pubyear{2018}
\date{Submitted 2018 July 31.}

\label{firstpage} 
\maketitle

\begin{abstract}
We analyzed the X-ray power density spectra of five ultraluminous X-ray sources (ULXs) \5408, \6946, \M82, \1313 and \IC342 that are the only ULXs which display both flat-topped noise (FTN) and quasi-periodic oscillations (QPO). We studied the QPO frequencies, fractional root-mean-square (rms) variability, X-ray luminosity and spectral hardness. We found that the level of FTN is anti-correlated with the QPO frequency. As the frequency of the QPO and brightness of the sources increase, their fractional variability decreases. 
We propose a simple interpretation using the spherizarion radius, viscosity time and $\alpha$-parameter as basic properties of these systems. The main physical driver of the observed variability is the mass accretion rate which varies ${\gtrsim}3$ between different observations of the same source. As the accretion rate decreases the spherization radius reduces and the FTN plus the QPO move toward higher frequencies resulting in a decrease of the fractional rms variability. We also propose that in all ULXs when the accretion rate is low enough (but still super-Eddington) the QPO and FTN disappear.  
Assuming that the maximum X-ray luminosity depends only on the black hole (BH) mass and not on the accretion rate (not considering the effects of either the inclination of the super-Eddington disc nor geometrical beaming of radiation) we estimate that all the ULXs have about similar BH masses, with the exception of \M82, which might be 10 times more massive.
\end{abstract}


\section{Introduction}
Ultraluminous X-ray sources (ULXs) are bright, point-like, extra-galactic sources with observed X-ray luminosities $L_X \ge 10^{39}$ erg s$^{-1}$ \citep[see][for a review]{Bachetti2016rev,Kaaret2017review}. 
The nature of these objects is still unclear. The luminosities of ULXs exceed the Eddington limit for a typical stellar-mass black hole (10\,M$_\odot$). Nevertheless they are not associated with galactic nuclei or background quasars. 
One of the possible models to explain their nature involves Intermediate Mass Black Holes (IMBHs) of $10^2$--$10^4$\,M$_\odot$ with standard accretion discs \citep{Colbert1999ULXIMBH}. This model implies that ULXs might be massive analogues of Galactic black hole binaries (BHBs). Another explanation is super-Eddington accretion onto stellar-mass black holes \citep{FabrikaMescheryakov2001,Poutanen2007}. A key feature of the super-Eddington accretion predicted by numerical simulations \citep{Ohsuga2005,Ohsuga2011} is a strong optically thick wind, which covers the inner parts of the disc and collimates the radiation. In this case spectral and variability properties of ULXs could be different from BHBs \citep{Gladstone2009} and resemble the superaccretor \SS433 in several aspects. In addition, the recent discovery of coherent X-ray pulsations in some sources \citep{Bachetti2014Nat,Furst2016pulsP13,Israel2017pulsP13, Israel2017pulsNGC5907} suggests that part of the whole population of ULXs are neutron stars accreting at super-Eddington rates \citep{King2001ULX,King2017PULX,MiddletonKing2017ULXbeaming}. 
       
\SS433 is the only known super-accretor in our Galaxy. It is a close binary system with a stellar-mass black hole (see \citealt{Fabrika2004} for a review). The accretion rate in this system is estimated to reach $300\dot{M}_{edd}$ (for the radiative efficiency of ${\eta}=1/12$ and black hole mass $M_{BH}\approx10M_{\sun}$, \citealt{Cherepashchuk2018SS433mass}), where $\dot{M}_{edd}$ is the Eddingon accretion rate. Optical spectra of \SS433 contain broad emission lines of hydrogen and He\,II with FWHM suggesting velocities of 1000 km/s originating from an accretion disc wind. The same line features have been revealed in the spectra from ULXs by \cite{Fabrika2015} from optical spectroscopy using the Subaru telescope. The latest showed that the optical spectra from ULXs are all very similar and resemble the optical spectrum from \SS433. This 
study shows that the lines present in the spectra support the idea of them arising from the wind funnel of a supercritical accretion disc instead from a donor-star wind or 
irradiated disc. Evidence of outflows have been suggested from X-ray spectral studies of \5408, \6946 \citep{Middleton2014wind} and \1313 \citep{Middleton2015atomic}. X-ray spectra from SS\,433 have been studied by \cite{FabrAbolmKarpov2007} applying a model based on the spectra from ULXs. The X-ray spectral residuals show a big bump around $\approx 1$\,keV due to the difference between the neutral ($N_H$) and highly ionised absorption outflows. More recently, blue-shifted absorption lines with velocities of $\sim$0.2c were directly measured by \cite{Pinto2016Nat} in the grating spectra of the same ULXs 
\citep[see also][]{Pinto2017NGC55ULX,Kosec2018NGC300ULX1}. These studies suggest the presence of highly ionized outflows in ULXs.

The X-ray spectra from ULXs have two common features: a soft excess below 2\,keV and a roll-over at higher energies. The latter was initially discovered in \XMM\ data \citep{Stobbart2006,Gladstone2009}. \textit{NuSTAR} observations revealed an unambiguous high-energy cutoff above 10\,keV \citep{Bachetti2013,Walton2014}. This spectral shape, hard to see in the case of BHBs, introduced the idea of a new, `ultraluminous' accretion state \citep{Roberts2007,Gladstone2009}. Depending on the ratio between the high- and low-energy parts of the spectrum, the ultraluminous state was splitted into three classes \citep{Sutton2013}: soft ultraluminous (SUL), hard ultraluminous (HUL) and broadened disc (BD). There is a trend for a particular source to belong to a specific class, although transitions between classes have also been observed \citep{Pintore2014}.

\SS433 is quite different from ULXs in X-rays. The apparent X-ray luminosity of this source is ${\sim}10^{36}$~erg s$^{-1}$ due to the high inclination of its accretion disc ($i\gtrsim 60^\circ$). Nevertheless the bolometric luminosity is much higher and has been estimated to be $\sim10^{40}$~erg~s$^{-1}$ \citep{Cherepashchuk2002,Fabrika2004,Middleton2018SS433}. The donor star in \SS433 contributes less than 15\,\% of the whole luminosity. The observed X-ray spectrum of \SS433 is mostly due to emission from jets and also by reflection of the radiation coming from innermost parts of the disc on the funnel walls \citep{MedvFabr2010}. It has been suggested that, if being observed near face-on, \SS433 might resemble a ULX.

Examining the fast X-ray variability of \SS433 \citep{Revn2006}, 
it has been found that its power density spectrum is well-fitted by a broken power-law with a flat component ($P\propto f^{0}$) below $10^{-3}$~Hz \citep{AtapinSS433var2015,Atapin2016pazh}. This type of variability is similar to the flat-topped noise (hereafter FTN) observed in BHBs at the state of low accretion rate, i.e. `low/hard' state (see \citealt{McClintockRemillard2006book,Belloni2016review} for reviews). However the standard states of black hole binaries do not correspond to what is typically observed in \SS433. Its spectral and variability behaviour depend on the precession phase of the system which is fully determined by the visibility of the funnel of the supercritical accretion disc. The flat component in the PDS of \SS433 only emerges when the funnel is maximally opened to the observer. This suggests that the FTN can be produced by supercritical accretion as well.

Some ULXs also show FTN often accompanied by quasi-periodic oscillations (QPO). This is most clearly seen in \5408 and \M82 which were the first ULXs where a QPO was discovered \citep{Strohmayer2003M82X1QPOdiscov,Strohmayer2007NGC5408QPOdiscov}. The fact that both QPO and FTN appear together and the much lower frequency of the former compared to what has been usually seen in BHBs was initially considered as an evidence for the presence of an IMBH \citep{Strohmayer2009NGC5408IMBH,Dheeraj2012NGC5408QPO,Pasham2013M82X1QPO}. Recent
studies using new data, in particular those regarding the discovery of the high-energy cutoff and the presence of an accelerated outflow, rule out the IMBH model, at least in the case of \5408. 
Also it has been shown that the PDS of \5408 does not appear to well match the type-C QPO observed in BHBs \citep{Middleton2011}. In the framework of super-Eddington accretion the observed variability properties of ULXs can be explained by the presence of an inhomogeneous wind \citep{Middleton2011,Sutton2013,Middleton2015model}.

QPOs have also been detected in \6946 \citep{Rao2010NGC6946QPOdiscov}, \IC342 \citep{Agrawal2015IC342QPOdiscov}, and in \1313 \citep{Pasham2015NGC1313X1QPOdiscov}. The QPO in Ho\,IX~X-1 reported by \citet{Dewangan2006HoIXQPOdiscov} has not been confirmed in later studies \citep{Heil2009}. Despite a significant sample of ULXs with QPO, their properties have never been studied together. We notice that both QPO and FTN may appear at the same dataset and in some other observations the QPO/FTN may disappear. They are indeed connected features, as we will show in this work. Indeed, the main goal of this paper is to perform a joined analysis of the timing data from these ULXs. This procedure  allows us to reveal any possible similarity in their variability properties and help us to understand the nature of the underlying physical processes driving it.

\section{Observations}
\begin{table*}
\begin{minipage}{165mm}
\caption{Log of the analyzed observations.}
\label{tab:data}
\begin{tabular}{lcccclcccc} \hline\hline
Obs. ID    & Date       & ${T_{obs}}^{a}$ & ${T_{exp}}^{b}$ & ${R_{net}}^{c}$ &
Obs. ID    & Date       & ${T_{obs}}^{a}$ & ${T_{exp}}^{b}$ & ${R_{net}}^{c}$ \\
           &            & (ks)            & (ks)            & (cnt/s)         &
           &            & (ks)            & (ks)            & (cnt/s)         \\ \hline\hline
\multicolumn{10}{|c|}{NGC\,5408~X-1 ($D=5.32$~Mpc$^d$)} \\ 
0302900101$^\dagger$ & 2006-01-13 & 132.2 & 90.0 & $0.538\pm0.005$ & 0653380401$^\dagger$ & 2011-01-26 & 121.0 & 91.8 & $0.615\pm0.005$    \\ 
0500750101$^\dagger$ & 2008-01-13 & 115.7 & 72.6 & $0.542\pm0.007$ & 0653380501$^\dagger$ & 2011-01-28 & 126.4 & 96.2 & $0.613\pm0.005$    \\
0653380201$^\dagger$ & 2010-07-17 & 128.9 & 76.5 & $0.691\pm0.005$ & 0723130301$^\dagger$ & 2014-02-11 & 38.5  & 34.3 & $0.564\pm0.009$    \\
0653380301$^\dagger$ & 2010-07-19 & 130.8 & 104.7& $0.672\pm0.005$ & 0723130401$^\dagger$ & 2014-02-13 & 36.0  & 32.9 & $0.599\pm0.010$    \\\hline
\multicolumn{10}{|c|}{NGC\,6946~X-1 ($D=6.72$~Mpc$^e$)} \\  
0200670301$^\dagger$ & 2004-06-13 & 15.6  & 8.7  & $0.333\pm0.014$  & 0500730101$^\dagger$ & 2007-11-08 & 31.9 & 25.7 & $0.306\pm0.010$    \\
0500730201$^\dagger$ & 2007-11-02 & 37.3  & 31.2 & $0.321\pm0.009$  & 0691570101$^\dagger$ & 2012-10-21 & 119.3& 97.4 & $0.386\pm0.003$    \\ \hline
\multicolumn{10}{|c|}{M82~X-1 ($D=3.7$~Mpc$^f$)} \\ 
0112290201$^\dagger$ & 2001-05-06 & 30.5 & 22.0 & $7.38\pm0.04$   & 0560590301 & 2009-04-29 & 52.9 & 16.4 & $8.69\pm0.07$ \\
0206080101$^\dagger$ & 2004-04-21 & 104.3& 56.1 & $5.89\pm0.03$ & 0657800101$^\dagger$ & 2011-03-18 &	26.6  & 14.8 & $6.26\pm0.03$ \\
0560590101 & 2008-10-03 & 31.9 & 25.5 & $13.63\pm0.17$ & 0657801901$^\dagger$ & 2011-04-29 & 28.2 & 8.2  & $5.53\pm0.04$ \\
0560181301 & 2009-04-03 & 27.3 & 9.2  & $7.92\pm0.03$  & 0657802101$^\dagger$ & 2011-09-24 & 22.8 & 8.4  & $5.84\pm0.03$ \\
0560590201 & 2009-04-17 & 44.6 & 12.5 & $11.55\pm0.12$ & 0657802301 & 2011-11-21 & 23.9 & 17.1 & $7.23\pm0.07$ \\ \hline
\multicolumn{10}{|c|}{NGC\,1313~X-1 ($D=3.88$~Mpc$^g$)} \\
0106860101 & 2000-10-17 & 42.4 & 15.7 & $0.991\pm0.008$ & 0693850501$^\dagger$ & 2012-12-16 & 125.2 & 96.4 & $1.141\pm0.004$ \\
0150280601 & 2004-01-08 & 53.3 & 6.3  & $1.693\pm0.022$ & 0693851201$^\dagger$ & 2012-12-22 & 125.2 & 86.3 & $1.173\pm0.005$ \\
0205230601 & 2005-02-07 & 14.3 & 9.9  & $1.109\pm0.016$ & 0722650101 & 2013-06-08 & 30.7 & 15.6 & $1.156\pm0.011$ \\
0405090101$^\dagger$ & 2006-10-15 & 123.2& 86.2 & $0.933\pm0.004$ & 0742590301 & 2014-07-05 & 63.0 & 60.0 &	$2.106\pm0.012$ \\ \hline
\multicolumn{10}{|c|}{IC\,342~X-1 ($D=3.93$~Mpc$^h$)} \\
0093640901 & 2001-02-11 & 10.9 & 5.9  & $0.738\pm0.015$ & 0206890401 & 2005-02-10 & 23.9  & 5.3  & $2.25\pm0.03$ \\ 
0206890101 & 2004-02-20 & 23.9 & 13.9 & $1.77\pm0.05$ & 0693850601$^\dagger$ & 2012-08-11 & 59.9 & 37.2 & $0.835\pm0.015$ \\  
0206890201 & 2004-08-17 & 23.9 & 19.6 & $0.862\pm0.007$  & 0693851301 & 2012-08-17 & 60.1  & 35.3 & $1.029\pm0.023$\\ \hline
\end{tabular}
\textit{Notes.} $^a$The total observation time. $^b$Useful exposure time after removing background flares. $^{c}$Net pn+MOS1+MOS2 count rate in the 1--10~keV band. $^d$Distance to the host galaxy is taken from \cite{Tully2013}. $^e$\cite{Tikhonov2014distNGC6946}. $^f$Median value of the distance from NED (http://ned.ipac.caltech.edu). $^g$\cite{Tikhonov2016distNGC1313}. $^h$\cite{Tikhonov2010distIC342}. $^\dagger$QPO was detected in this observation.
\end{minipage}
\end{table*}

In this paper we analyse the short-term variability of five ULXs: \5408, \6946, \M82, \1313 and \IC342. 
We used data from the \XMM\ archive. All public observations longer than 10~ks have been included, excluding data sets affected by strong background flaring. We also excluded some observations of \1313 where the source was positioned on the chip gap (this constitutes $1/4$ of the whole dataset only). All the analysed data sets are listed in Table~\ref{tab:data}.

The observations were reduced using the {\it XMM-Newton} \textsc{sas} version 15.0.0. We used the data from the European photon imaging camera (EPIC) pn and both MOS detectors. The first four observations of \IC342 were taken in `extended full-frame' mode of the pn detector. All the other observations in Table~\ref{tab:data} are in `full-frame' imaging mode (time resolution is 73.4~ms and 2.6~s for pn and MOS\footnote{Resolution of the MOS1 has been reduced to 2.7\,s for observations after December 2012:\\https://www.cosmos.esa.int/web/xmm-newton/mos1-ccd3} respectively). The data sets were filtered with patterns 0--4 for pn and 0--12 for MOS including only single and double pixel events and applying `FLAG==0’ to exclude the events from bad pixels and events near chip gaps.

Particular attention was taken removing background flares because of the spurious variability introduced by flaring may distort the power spectra. First of all, we extracted pn light curves from the entire field of view in the 10--12~keV band. The count rate of non-flaring background have proven to be slightly different between the observations. For most data sets we applied the standard screening criterion\footnote{https://www.cosmos.esa.int/web/xmm-newton/sas-thread-epic-filterbackground} including only time periods where the pn count rate was not higher than 0.4 count s$^{-1}$. However we found that for some observations with stable background this threshold rejects too many good data points. In these cases we used a less restrictive condition and included time periods where the count rate was not higher than three times the standard deviations from the mean.

The flux from \M82 is significantly contaminated by diffuse emission from the host galaxy and by another bright ULX~--- M82~X-2. The separation between X-1 and X-2 is only 5’’ \citep{Matsumoto2001M82}, which is near the limit of the \XMM\ spatial resolution. Both ULXs are highly variable and have about similar luminosities. We were not able to avoid this contamination, however, below we discuss how it can affect the results of our analysis (Sec.~3.2).

\begin{table*}
\begin{minipage}{175mm}
\caption{Best fitting model parameters for the PDS in which we detected QPO.}  
\label{tab:pds_fit}
\begin{tabular}{lcccccccccc} \hline\hline
Obs. ID & ${f_q}^a$ & ${W_q}^b$ & ${Q}^c$ & ${R_q}^d$ & ${f_b}^e$ & ${\beta}^f$ & $A^g$ & $\chi^2$/dof & ${S_b}^h$ & ${S_q}^i$\\ 
&  mHz      & mHz     &  & (rms/mean)$^2$ & mHz   &     &   &            & $\sigma$  & $\sigma$ \\ \hline \hline
\multicolumn{9}{|c|}{NGC\,5408~X-1} \\          
0500750101 & $10.9^{+1.5}_{-2.9}$ & $9^{+6}_{-4}$ & 1.2 & $10^{+9}_{-5}\cdot 10^{-2}$ & 
$4.1^{+2.4}_{-1.6}$ & $>1.35$ & $19^{+5}_{-3}$ & 231.7/201 & 5.2 & 3.4 \\
0723130301 & $11.9^{+2.1}_{-1.7}$ & $10^{+3}_{-4}$ & 1.2 & $12.9^{+2.4}_{-7.1}\cdot 10^{-2}$ & 
$4.8^{+1.6}_{-1.3}$ & $>1.64$ & $15^{+4}_{-3}$ & 230.6/210 & 5.0 & 3.3  \\
0723130401 & $13.9^{+1.1}_{-0.9}$ & $5^{+5}_{-3}$ & 2.8 & $6^{+4}_{-4}\cdot 10^{-2}$ & 
$7^{+6}_{-3}$ & $1.54^{+0.7}_{-0.4}$ & $11.7^{+2.6}_{-2.2}$ & 142.2/162 & 5.3 & 3.0 \\
0302900101 & $18.1^{+2.1}_{-2.0}$ & $20^{+5}_{-6}$ & 0.9 & $13.3^{+2.4}_{-5.9}\cdot 10^{-2}$ & 
$5.9^{+1.5}_{-1.9}$ & $>1.26$ & $7.8^{+2.4}_{-1.5}$ & 152.0/155 & 4.9 & 5.1 \\
0653380501 & $18.2^{+2.3}_{-3.1}$ & $19^{+6}_{-7}$ & 1.0 & $11^{+5}_{-5}\cdot 10^{-2}$ & 
$6.5^{+2.7}_{-2.0}$ & $>1.29$ & $7.3^{+1.8}_{-1.7}$ & 117.2/113 & 5.1 & 4.1 \\
0653380401 & $18.7^{+1.9}_{-2.2}$ & $20^{+4}_{-5}$ & 0.9 & $10.8^{+2.7}_{-3.5}\cdot 10^{-2}$ & 
$5.6^{+2.3}_{-1.7}$ & $>1.48$ & $5.6^{+1.2}_{-1.2}$ & 172.1/169 & 4.3 & 4.0 \\
0653380301 & $39.3^{+2.7}_{-3.2}$ & $22^{+15}_{-11}$ & 1.8 & $3.6^{+3.5}_{-2.1}\cdot 10^{-2}$ & 
$18^{+7}_{-7}$ & $>1.02$ & $2.5^{+0.6}_{-0.5}$ & 69.8/63 & 3.7 & 3.0 \\ 
0653380201 & $41.8^{+2.8}_{-9.8}$ & $22^{+27}_{-16}$ & 1.9 & $3.3^{+6.4}_{-2.4}\cdot 10^{-2}$ &  
$19^{+10}_{-8}$ & $>0.96$ & $2.5^{+0.7}_{-0.7}$ & 59.2/73 & 3.7 & 3.0 \\ \hline
\multicolumn{9}{|c|}{NGC\,6946~X-1} \\  
0500730101 & $8.7^{+0.8}_{-0.9}$ & $1.6^{+2.2}_{-1.1}$ & 5.4 & $6^{+8}_{-4}\cdot 10^{-2}$ & 
$3.4^{+1.4}_{-1.2}$ & $1.58^{+0.28}_{-0.25}$ & $66^{+20}_{-16}$ & 326.4/247 & 3.5 & 2.8 \\
0500730201 & $9.4^{+1.1}_{-0.9}$ & $5.2^{+3.4}_{-2.2}$ & 1.8 & $14^{+9}_{-6}\cdot 10^{-2}$ & 
$2.2^{+1.6}_{-1.3}$ & $1.26^{+0.6}_{-0.3}$ & $45^{+24}_{-14}$ & 286.8/292 & 3.2 & 3.3 \\
0200670301 & $27^{+5}_{-4}$ & $5.2^{+3.4}_{-1.3}$ & 5.2 & $6^{+4}_{-4}\cdot 10^{-2}$ & 
$12.0^{+2.7}_{-3.7}$ & $>1.33$ & $11^{+4}_{-3}$ & 80.0/77 & 3.6 & 2.6 \\
0691570101 & $41.7^{+2.4}_{-3.7}$ & $23^{+16}_{-10}$ & 1.8 & $9^{+8}_{-5}\cdot 10^{-2}$ & 
$16^{+9}_{-5}$ & $>0.86$ & $4.1^{+1.2}_{-1.0}$ & 78.0/93 & 6.1 & 3.4 \\  \hline
\multicolumn{9}{|c|}{M82~X-1} \\ 
0657802101 & $35.2^{+1.9}_{-2.4}$ & $7.9^{+4.7}_{-2.8}$ & 4.5 & $6.0^{+2.8}_{-2.5}\cdot 10^{-3}$ & 
$7^{+3}_{-3}$ & $1.42^{+1.2}_{-0.7}$ & $0.42^{+0.14}_{-0.14}$ & 97.4/94 & 2.6 (3.7) &3.3 \\
0657801901 & $46.2^{+1.9}_{-3.0}$ & $9^{+6}_{-5}$ & 5.1 & $9^{+4}_{-4}\cdot 10^{-3}$ & 
$27^{+20}_{-14}$ & $>1.18$ & $0.15^{+0.7}_{-0.7}$ & 84.2/75 & 2.9 (3.2) & 3.5 \\
0657800101 & $47.0^{+2.2}_{-2.3}$ & $8^{+7}_{-5}$ & 5.8 & $4.5^{+2.4}_{-2.2}\cdot 10^{-3}$ & 
$11^{+5}_{-5}$ & $>0.89$ & $0.27^{+0.26}_{-0.10}$ & 55.0/42 & 2.2 (3.5) & 2.9 \\
0112290201 & $52.7^{+3.7}_{-2.6}$ & $15^{+11}_{-7}$ & 3.5 & $3.6^{+1.8}_{-1.4}\cdot 10^{-3}$ & 
$8.9^{+12}_{-5}$ & $>1.13$ & $0.22^{+0.11}_{-0.08}$ & 70.7/70 & 2.7 (4.6) & 3.9 \\
0206080101 & $119^{+3}_{-5}$ & $26^{+19}_{-12}$ & 4.57 & $9.9^{+5}_{-2.9}\cdot 10^{-3}$ & 
$110^{+90}_{-80}$ & $>0.35$ & $0.039^{+0.023}_{-0.025}$ & 66.3/82 & 1.3 (2.5) & 6.2 \\ \hline
\multicolumn{9}{|c|}{NGC\,1313~X-1} \\
0405090101 & $80.4^{+2.9}_{-2.2}$ & $10^{+6}_{-5}$ & 8.0 & $1.7^{+0.9}_{-1.1}\cdot 10^{-2}$ & 
$60^{+40}_{-30}$ & $>0.96$ & $0.28^{+0.16}_{-0.07}$ & 133.2/131 & 1.3 (2.4) &3.2 \\ 
0693851201 & $268^{+21}_{-15}$ & $70^{+65}_{-50}$ & 3.8 & $4.2^{+3.0}_{-2.2}\cdot 10^{-2}$ & 
$<130$ & $>0.65$ & $0.27^{+0.41}_{-0.09}$ & 40.0/55 & 1.3 (2.3) & 3.6  \\ 
\multirow{2}{*}{0693850501$^j$} & $297^{+5}_{-6}$ & $24^{+30}_{-22}$ & 12.4 & $2.9^{+1.4}_{-1.5}\cdot 10^{-2}$ &  
\multirow{2}{*}{$<350$} & \multirow{2}{*}{$>0.74$} & \multirow{2}{*}{$0.11^{+0.08}_{-0.06}$} & \multirow{2}{*}{54.5/66} & \multirow{2}{*}{1.2 (2.1)} & 4.4 \\ 
& $630^{+60}_{-40}$ & $80^{+100}_{-50}$ & 7.8 & $5.3^{+3.1}_{-2.8}\cdot 10^{-2}$ & 
& & & & & 2.7 \\ \hline
\multicolumn{9}{|c|}{IC\,342~X-1} \\
0693850601 & $654^{+26}_{-30}$ & $80^{+70}_{-50}$ & 8.2 & $7^{+11}_{-3}\cdot 10^{-2}$ & 
$<320$ & $>0.51$ & $<0.29$ & 43.3/73 & $<1$ (1.2) & 3.2 \\  \hline     
\end{tabular}
\textit{Note.} The data sets are ordered by the centroid frequency of the QPO. All uncertainties refer to 90\% confidence level. $^a$Centroid frequency of the QPO. $^b$FWHM of the QPO. $^c$Quality factor of the QPO $Q=f_q/W_q$. $^d$Total power under the QPO profile. $^e$Break frequency of the flat-topped noise (FTN). $^f$Power-law index above the break. $^g$Level of the FTN (in units of (rms/mean)$^2$ Hz$^{-1}$). $^h$Statistical significance of the flattening of power spectrum at lower frequencies and also significance of the presence of any power besides QPO above the Poisson noise (shown in brackets) obtained  via F-test. $^i$Statistical significance of the QPO obtained via F-test. $^j$Two Lorentzians were used to fit this observation.
\end{minipage}
\end{table*}

\begin{figure*}
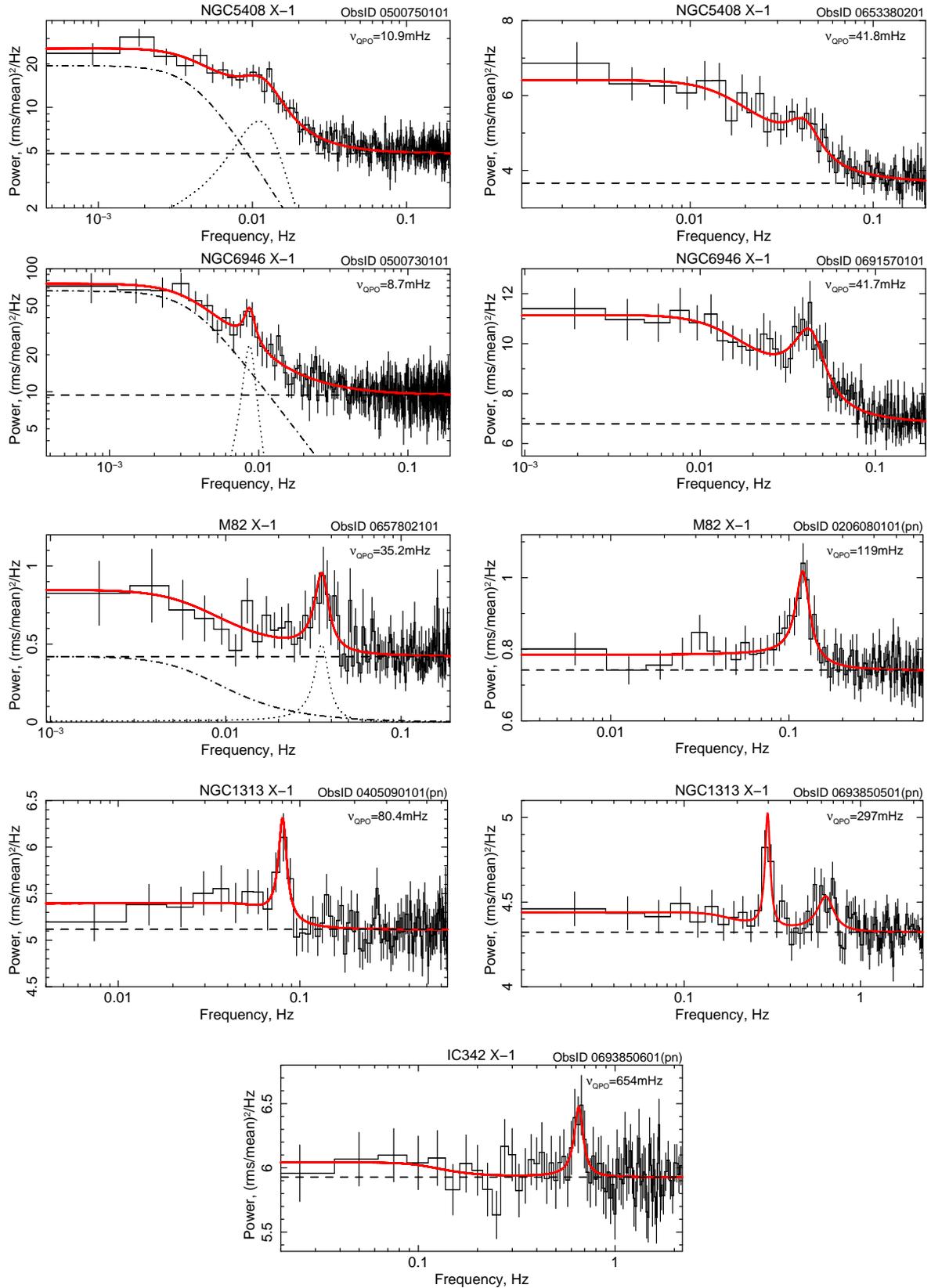

\includegraphics[angle=-90,width=0.45\textwidth]{fig1_11}
\includegraphics[angle=-90,width=0.45\textwidth]{fig1_12}
\vspace*{3mm}
\includegraphics[angle=-90,width=0.45\textwidth]{fig1_21}
\includegraphics[angle=-90,width=0.45\textwidth]{fig1_22}
\vspace*{3mm}
\includegraphics[angle=-90,width=0.45\textwidth]{fig1_31}
\includegraphics[angle=-90,width=0.45\textwidth]{fig1_32}
\vspace*{3mm}
\includegraphics[angle=-90,width=0.45\textwidth]{fig1_41}
\includegraphics[angle=-90,width=0.45\textwidth]{fig1_42}
\vspace*{3mm}
\centering{\includegraphics[angle=-90,width=0.45\textwidth]{fig1_5}}
\vspace*{-3mm}
\caption{Power density spectra of five ULXs that were reported to exhibit QPO (energy band of 1--10~keV). The observations with the lowest and the highest QPO frequency for each source are shown on the left-hand and right-hand sides respectively. The level of Poisson noise is shown by dashed lines. The solid lines are the best-fitting models, the dotted and dash-doted lines represent model components: Lorentzians and broken power law (these components have much lower levels than Poisson noise in most cases and are well below the lower boundaries of the plots). It is clearly seen that a lower QPO centroid frequency corresponds to a higher power in the PDS above Poisson noise.}
	\label{fig:pds}
\end{figure*}

Source light curves were extracted from circular regions with a radius of 32’’. For \M82 we additionally used smaller aperture of 12’’ radius to reduce the contribution of the emission from the host galaxy, and compared the results obtained from both apertures. For all the sources except \M82 which cannot be resolved as a separate point source, the circle position was adjusted to be centred around the brightest pixel in each observation. In the case of \M82 the centre of the region was fixed at RA = 09 55 50.2, Dec = +69 40 47 \citep{Matsumoto2001M82}. Background regions were chosen to be as large as possible to diminish statistical fluctuations, located in the same chip of the source in areas free of other sources and away from the read-out direction. Light curves were extracted with the task \textsc{evselect} with the same bin-size for all the detectors. To prevent possible asynchronicity between pn and MOS light curves, we launched this task with parameters `timemin’ and `timemax’ \citep{BarnardTrudolyubov2007} assigning the same start and stop time for each detector. Background light curves were subtracted using the task \textsc{epiclccorr}, which takes into account instrumental effects like vignetting, bad pixels, etc. Then we created combined pn+MOS1+MOS2 light curves for all the observations. 

\section{Timing analysis}
\subsection{Power spectra}
Studies of the power spectra of \5408 \citep{Dheeraj2012NGC5408QPO,CaballeroGarcia2013NGC5408X1} and \M82 \citep{Pasham2013M82X1QPO,CaballeroGarcia2013M82X1QPO} have shown that the properties of the QPO vary between different observations. Sometimes in \M82 the QPO disappears. 
So, first of all, we examined all the data sets to check for the presence of the QPO and make a preliminary estimate of its centroid frequency. In order to increase the signal we
started producing the PDS from combined pn+MOS light curves that covered frequencies up to the Nyquist frequency of $f_{nq}\approx0.2$~Hz. If the QPO peak was absent in this frequency range, we produced a pn-only PDS, which allowed to extend the frequency range up to $f_{nq}\approx6.8$~Hz. Eventually, QPOs have been detected in 21 data sets of 36 and are listed in Table~\ref{tab:data}. We have analyzed all the observations with QPO from \5408 and \6946, five observations from \M82, three from \1313 and only one observation from \IC342. We explored the PDS of these observations in more detail. However the failure to detect QPO in the rest of PDS might still be caused by insufficient length of those observations.

\begin{figure*}
\includegraphics[angle=0,width=0.45\textwidth]{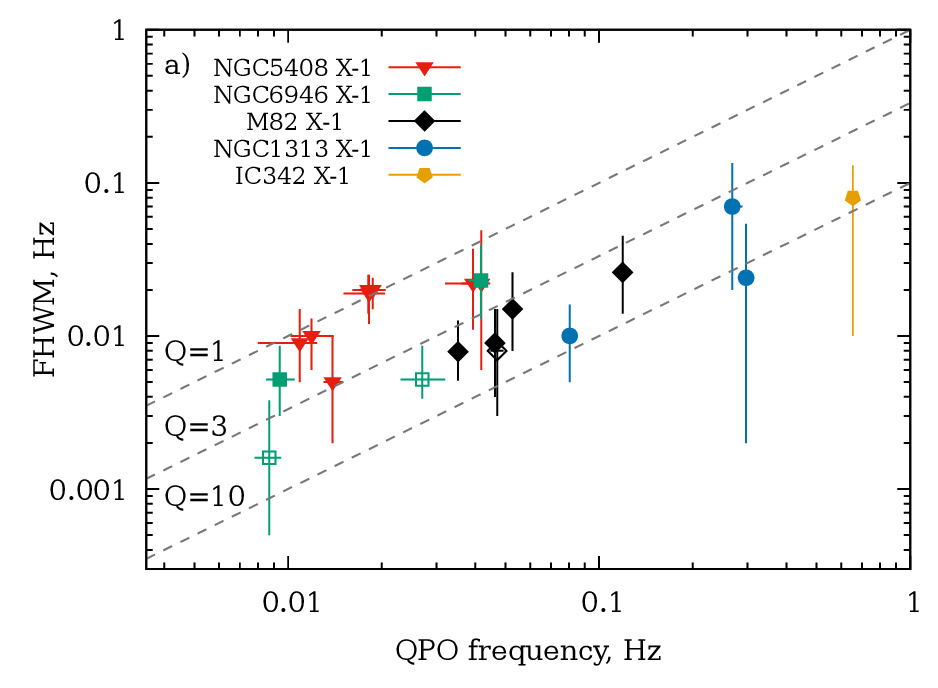}
\includegraphics[angle=0,width=0.45\textwidth]{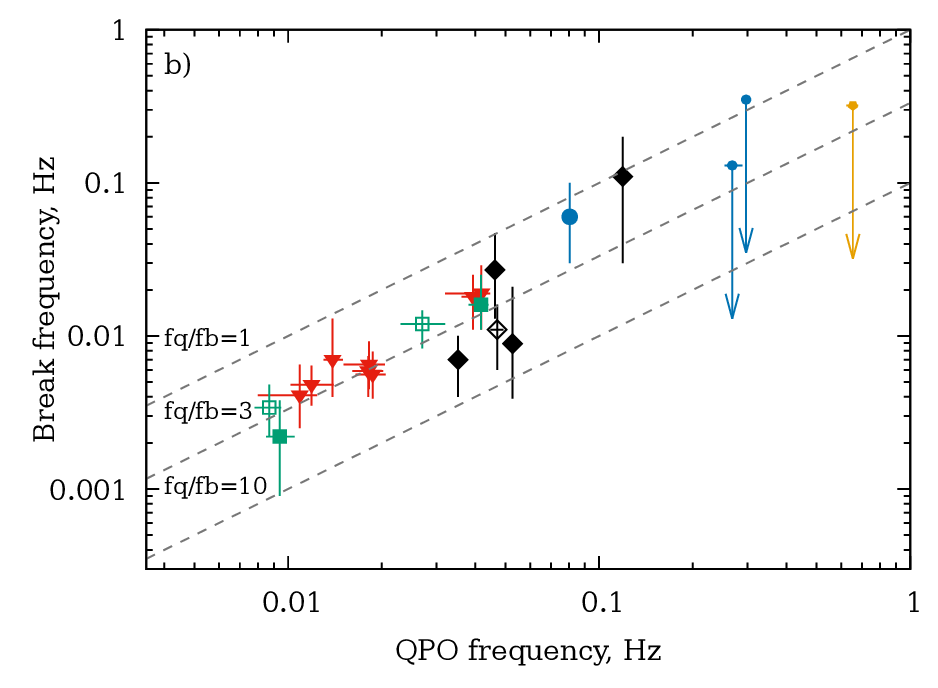}
\caption{\textit{Left panel}: The FWHM of the QPO as a function of its centroid frequency. The dashed lines correspond to constant quality factors $Q=f_q/W_q=1$, 3 and 10. \textit{Right panel}: The QPO centroid frequency versus the break frequency of the flat-topped noise. Arrows denote observations where we obtained only the upper limit of $f_b$ (Table~\ref{tab:pds_fit}). The dashed lines represent ratios $f_q/f_b=1$, 3 and 10. We found mean $f_q/f_b\approx 3$. Unfilled data points denote QPO with significance less than $3\sigma$ (observations of \6946 and \M82).}
\label{fig:fq_fwhm}
\end{figure*}

The PDS were constructed as follows. Since the fast X-ray variability of ULXs is stronger at higher energies \citep{Middleton2011,DeMarco2013}, we used light curves in the energy band of 1--10~keV. To increase the signal-to-noise ratio, we divided the light curves into shorter segments of equal length, produced individual power spectra from each segment and averaged them out. The length of the segments was chosen separately for each observation to make the lowest frequency in the resultant PDS be around one tenth~-- one fiftieth of the QPO frequency. Some segments had gaps resulting from the background flaring removing and telemetry dropouts. We rejected such segments when the fraction of the gaps was greater than 5\% of the length of the segments. Otherwise the gaps were filled with values of the count rates having a normal distribution with the same mean and rms as the original time series. Then the PDS were normalized to the fractional rms squared normalization \citep{vanderKlis1997,Vaughan2003}. The PDS with the lowest and the highest QPO frequencies for each source are shown in Fig.~\ref{fig:pds}.

All the PDS in which we detected QPO have a similar structure. The power increases towards lower frequencies as a power law and bends to the flat spectrum  $P\propto f^0$ (FTN) near the break frequency $f_b$. The QPO peak is located on the power-law part of the PDS close to the break. At the highest frequencies the power spectrum is dominated by Poisson noise.

The PDS were fitted using \textsc{xspec} version 12.9.0i with a model consisting of three components. The QPO peak was approximated by a Lorentzian with a centroid frequency $f_{q}$, FWHM width $W_q$ and the total area under the Lorentzian $R_q$.
To describe the continuum we used a broken power law
\begin{equation}
P \propto \frac{A}{f^{\beta_0} \sqrt{1+(f/f_b)^{2\beta}}},
\label{eq:pdsbkn}
\end{equation}
with index $\beta$ above the break and $\beta_0$ fixed at zero to represent the white noise below the break. 
The normalization of this model was chosen so that $A$ was the level of the flat component. Poisson noise was modelled by a constant which was allowed to vary. The obtained model parameters are shown in Table~\ref{tab:pds_fit}. 

For each PDS, we estimated the statistical significance of the QPO using the F-test. However, it has been noted that the classical F-test may give an incorrect result in the case of testing for the presence of additive model components like QPO in PDS or emission lines in energy spectra \citep{Protassov2002} because the test statistic may not follow the Fisher distribution. Additional procedures are needed to calibrate the test statistic. Therefore, for each data set we performed simulations as proposed by \cite{Protassov2002}. First of all, we fitted each observed PDS by two models: the model described above (target model) and the continuum-only model (null model). All the model parameters (including the QPO and break frequencies) were allowed to vary. For both models, we obtained the $\chi^2$ and computed the \textit{observed F-statistic} via standard formulae accounting for the difference in the number of degrees of freedom between the two models. Then for each data set we simulated 10\,000 PDS according to the null model with parameters fixed at their best-fit values using the \textsc{xspec} command `fakeit' ( to obtain the Gaussian errors in the simulated PDS we made sure that the POISSERR keyword in the FITS header is set to "false") and fitted them in the same way to obtain the \textit{simulated F-statistic}. Comparing  the F-statistics for the simulated data with the observed one, we obtained the p-value (probability that the observed PDS could be properly fitted without the Lorentzian model component). Eventually, we have found that the p-values derived from simulations and inferred from the Fisher distribution (classical F-test) are consistent with each other within 30\% in all the cases\footnote{Due to the limited sample, the simulations cannot reveal p-values less than $1\cdot10^{-4}$ ($\sigma\approx 3.9$).}. For clarity, in Table~\ref{tab:pds_fit} the p-values are converted to sigmas.

Similar simulations were carried out to test the significance of the FTN. As a null model we used one consisting of a power law (without break), Lorentzian and a constant for the Poisson noise. The probabilities obtained via F-test (Table~\ref{tab:pds_fit}) suggest that only \5408 and \6946 show strong evidence of a break. The significances of the breaks in \M82, \1313 and \IC342 are much lower. This is largely due to the strong Poisson noise which dominates over the intrinsic variability of these sources and does not allow to constrain the shape of their PDS continua. Thus, for these sources we additionally tested the presence of any variability present in the continuum. In this case, as a null model we used 'Lorentzian+constant` and performed simulations similar to those described above. The obtained probabilities converted to sigmas are provided in the same column of Table~\ref{tab:pds_fit} but in brackets.

Our measurements of the QPO centroid frequencies are consistent with the results obtained by other authors (\5408~-- \citealt{Dheeraj2012NGC5408QPO,CaballeroGarcia2013NGC5408X1,DeMarco2013}; \6946~-- \citealt{Rao2010NGC6946QPOdiscov}; \M82~-- \citealt{Pasham2013M82X1QPO,CaballeroGarcia2013M82X1QPO}; \1313~-- \citealt{Pasham2015NGC1313X1QPOdiscov}; \IC342~-- \citealt{Agrawal2015IC342QPOdiscov}) for all the observations except ObsID~0657802301 of \M82. We did not detect any QPO in this observation whilst \cite{Pasham2013M82X1QPO} reported one at $f_q\approx 0.2$~Hz. Also we have not found evidence for a second highly significant QPO peak in any observation except ObsID~0693850501 (\1313). In the latter case, we placed the second Lorentzian near 0.6~Hz (Fig.~\ref{fig:pds}) improving the $\chi^2$ from 65.4/69 to 54.5/66, however, this improvement is still insufficient to consider it as a firm detection ($\sigma\approx 2.7$). For ObsID~0657800101 (\M82) we obtained the QPO significance $\sigma\approx 2.9$. Nevertheless, we included this observation in further analysis because it has already been studied by \cite{Pasham2013M82X1QPO} and \cite{CaballeroGarcia2013M82X1QPO} who estimated its significance as 3.9 and 3.6, respectively. We also included two observations of \6946: ObsID~0500730101 \cite[analysed by][]{Rao2010NGC6946QPOdiscov} and ObsID~0200670301, obtaining $\sigma\approx 2.8$ and 2.6, respectively. This level should be considered only as a marginal detection, therefore we marked these data sets in the figures below. 

For \M82 we produced PDS taken from 32’’ and 12’’ radius apertures and found that the QPO properties are similar in both cases. In Table~\ref{tab:pds_fit} we prefer to present the model parameters from the PDS for the 32’’ -radius because they are more stable and have lower uncertainties. The spectral index $\beta$ turned out to be poorly constrained for most of the observations, since the power-law portion in the PDS near the QPO is better described by the wing of the Lorentzian than by the broken power law. We therefore provide only lower limits of $\beta$ for most of the PDS in Table~\ref{tab:pds_fit}.

The QPO frequencies, considering the sample of five sources as a whole, vary in a broad range spanning about two orders of magnitude. The lowest frequency is seen in \6946 (9.4~mHz, or 8.7~mHz taking into account ObsID~0500730101 with marginally detected QPO) and the highest in \IC342 (654~mHz). However, each source shows variations of $f_q$ by a factor of 3--5. In Fig.~\ref{fig:fq_fwhm}a we plot the QPO centroid frequency versus the FWHM of the QPO peak. The quality factors of the QPO ($Q=f_q/W_q$) in \5408 and in two observations of \6946 are about unity, which is lower than usually assumed for QPO ($Q\gtrsim2$). Other authors reported similar values (within errors) in these observations \citep{Pasham2013M82X1QPO,DeMarco2013,Rao2010NGC6946QPOdiscov}. The rest of ULXs showed higher quality factors ($Q\sim 3-10$).

The break in the PDS moves together with the QPO peak frequency between different observations keeping the ratio $f_q/f_b$ nearly constant (within uncertainties). For \5408, \6946 and \M82 we obtained the mean $f_q/f_b\approx 3$ (Fig.~\ref{fig:fq_fwhm}b, however, in the case of \M82 the shape of the continuum is not well-constrained). In sub-Eddington BHBs this ratio is about 10 \citep{Wijnands1999QPObreak}, i.e. the QPO peak is located much farther from the break. It has been noted by \cite{Middleton2011} that such a discrepancy may suggest a difference in the accretion states between BHBs and ULXs.

\subsection{Correlations between the QPO frequency, fractional variability and luminosity}
\cite{DeMarco2013} have revealed a relation between the centroid frequency of the QPO and variability of the source when analyzing six long observations of \5408. The total amount of power in the PDS decreases as the QPO frequency increases. We found that the same is true for all the five ULXs from our sample (Fig.~\ref{fig:pds}). The anti-correlation between the FTN level and the QPO centroid frequency is seen in Table~\ref{tab:pds_fit} (parameters $A$ and $f_q$ respectively, at least for \5408, \6946 and \M82 where the flat components are highly significant). Parameter $A$ has a low accuracy due to high uncertainties of the PDS at the lowest frequencies.

\begin{table}
\begin{minipage}{85mm}
\caption{Fractional rms variability (including contribution from both FTN and QPO) in the 0.001--0.2~Hz frequency range.}
\label{tab:frms}
\begin{tabular}{lclc} \hline\hline
Obs. ID    & ${F_{rms}}$ & 
Obs. ID    & ${F_{rms}}$ \\
&            & 
&            \\ \hline\hline
\multicolumn{4}{|c|}{NGC\,5408~X-1} \\ 
0302900101$^\dagger$ & $0.445\pm 0.009$ & 0653380401$^\dagger$ & $0.404\pm 0.010$   \\ 
0500750101$^\dagger$ & $0.489\pm 0.014$ & 0653380501$^\dagger$ & $0.405\pm 0.008$   \\
0653380201$^\dagger$ & $0.317\pm 0.012$ & 0723130301$^\dagger$ & $0.454\pm 0.017$	\\
0653380301$^\dagger$ & $0.329\pm 0.010$ & 0723130401$^\dagger$ & $0.451\pm 0.015$   \\ \hline
\multicolumn{4}{|c|}{NGC\,6946~X-1} \\  
0200670301$^\dagger$ & $0.43\pm 0.07$   & 0500730101$^\dagger$ & $0.65\pm 0.04$  \\
0500730201$^\dagger$ & $0.63\pm 0.04$   & 0691570101$^\dagger$ & $0.412\pm0.016$ \\ \hline
\multicolumn{4}{|c|}{M82~X-1} \\ 
0112290201$^\dagger$ & $0.063\pm 0.008$ & 0560590301 & $0.034\pm 0.008$             \\
0206080101$^\dagger$ & $0.089\pm 0.005$ & 0657800101$^\dagger$ & $0.098\pm 0.006$ \\
0560590101 &  $0.015\pm 0.008$          & 0657801901$^\dagger$ & $0.107\pm 0.016$ \\
0560181301 &  $0.040\pm 0.012$          & 0657802101$^\dagger$ & $0.110\pm 0.011$ \\
0560590201 &  $0.025\pm 0.009$          & 0657802301 & $0.036\pm 0.009$           \\ \hline
\multicolumn{4}{|c|}{NGC\,1313~X-1} \\
0106860101 & $0.123\pm 0.023$           & 0693850501$^\dagger$ & $0.093\pm 0.011$ \\
0150280601 & $0.051\pm 0.026$           & 0693851201$^\dagger$ & $0.100\pm 0.010$ \\
0205230601 & $0.13\pm 0.04$           & 0722650101 & $0.12\pm 0.03$           \\
0405090101$^\dagger$ & $0.148\pm 0.012$ & 0742590301 & $0.035\pm 0.011$           \\ \hline
\multicolumn{4}{|c|}{IC\,342~X-1} \\
0093640901 & $0.13\pm0.05$            & 0206890401 & $0.029\pm0.022$           \\ 
0206890101 & $0.06\pm0.03$            & 0693850601$^\dagger$ & $0.082\pm 0.016$ \\  
0206890201 & $0.069\pm0.016$          & 0693851301 & $0.06\pm0.03$           \\ \hline
\end{tabular}
\textit{Note.} The data sets are ordered in the same way as in Table~\ref{tab:data}. $^\dagger$QPO was detected in this observation.
\end{minipage}
\end{table} 

\begin{figure}
\includegraphics[angle=0,width=0.45\textwidth]{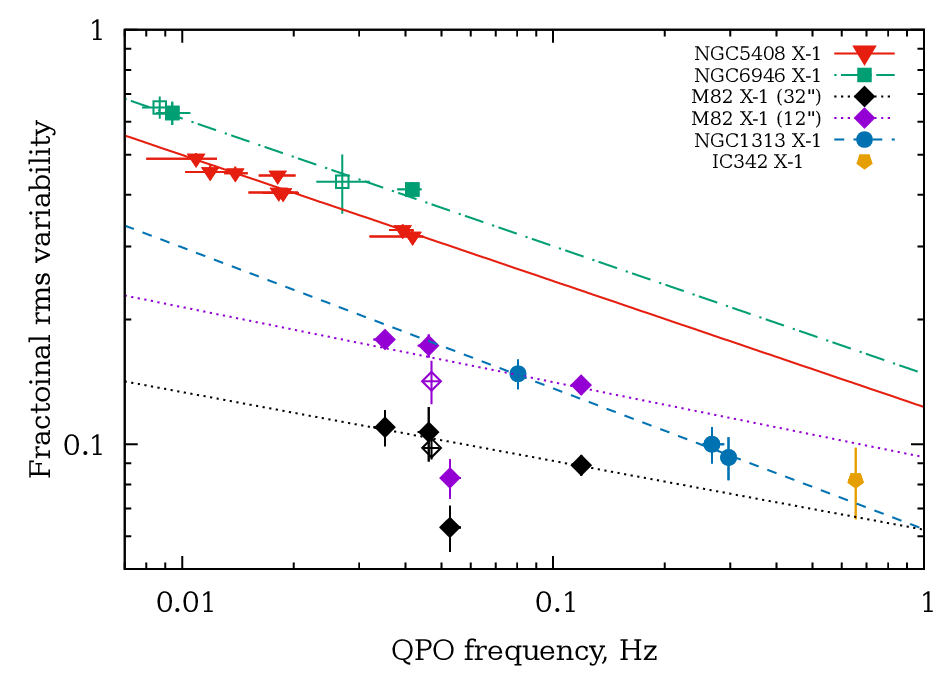}
\caption{Fractional rms variability in the frequency range of $0.001-0.2$~Hz in the 1--10~keV energy band as a function of the best fit QPO centroid frequency. For \M82 we show the fractional rms variability taken from both 32'' and 12'' radius apertures. Lines of different types represent the power-law fits to the data point of different source. Unfilled data points denote QPO with significance less than $3\sigma$ (Table~\ref{tab:pds_fit}).}
\label{fig:qpo_rms}
\end{figure}

\begin{figure}
\includegraphics[angle=0,width=0.45\textwidth]{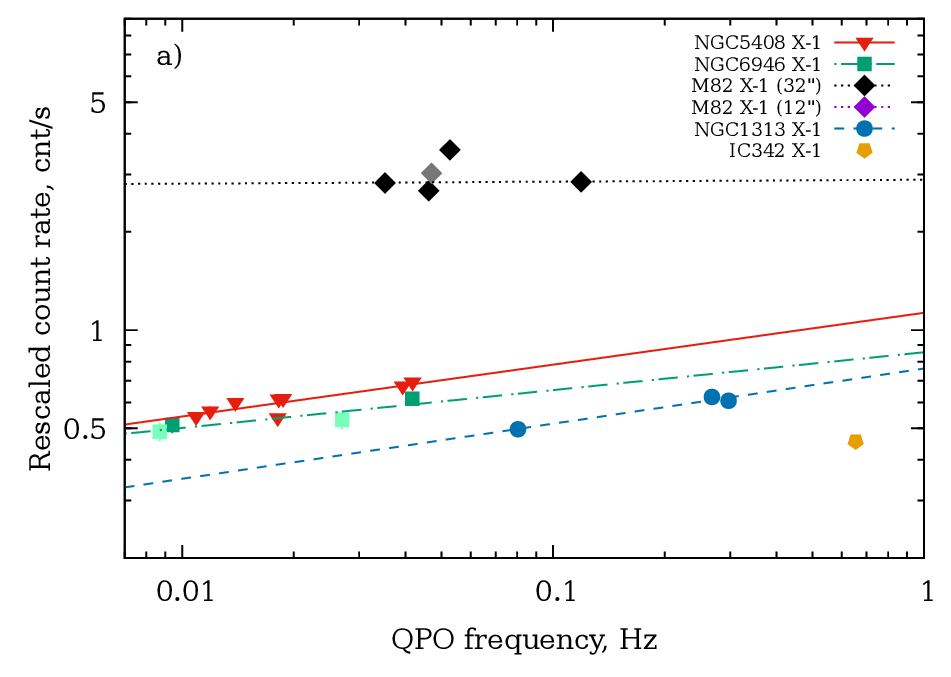}
\includegraphics[angle=0,width=0.45\textwidth]{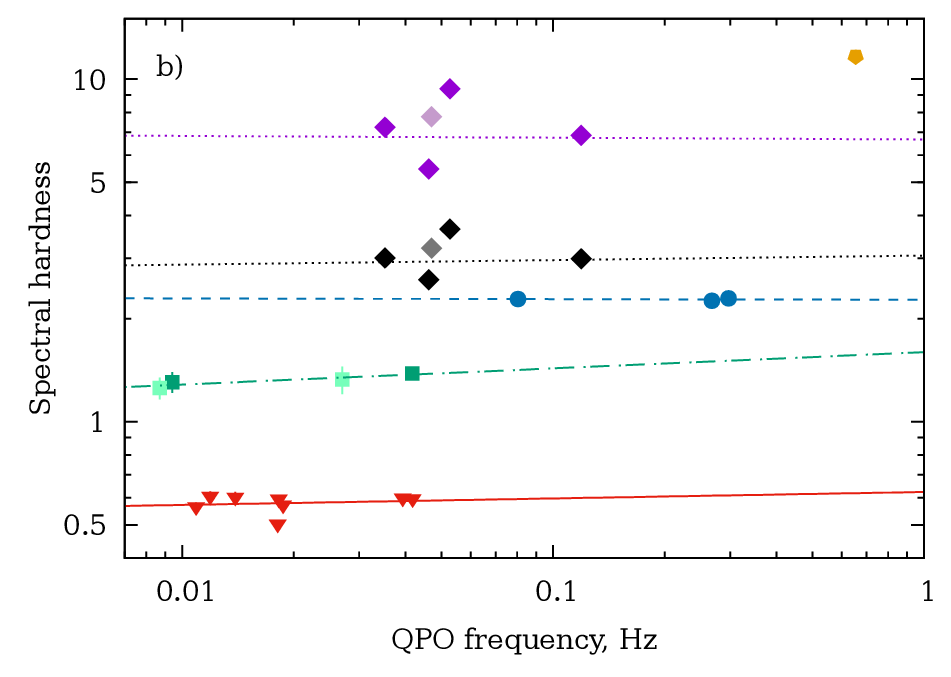}
\includegraphics[angle=0,width=0.45\textwidth]{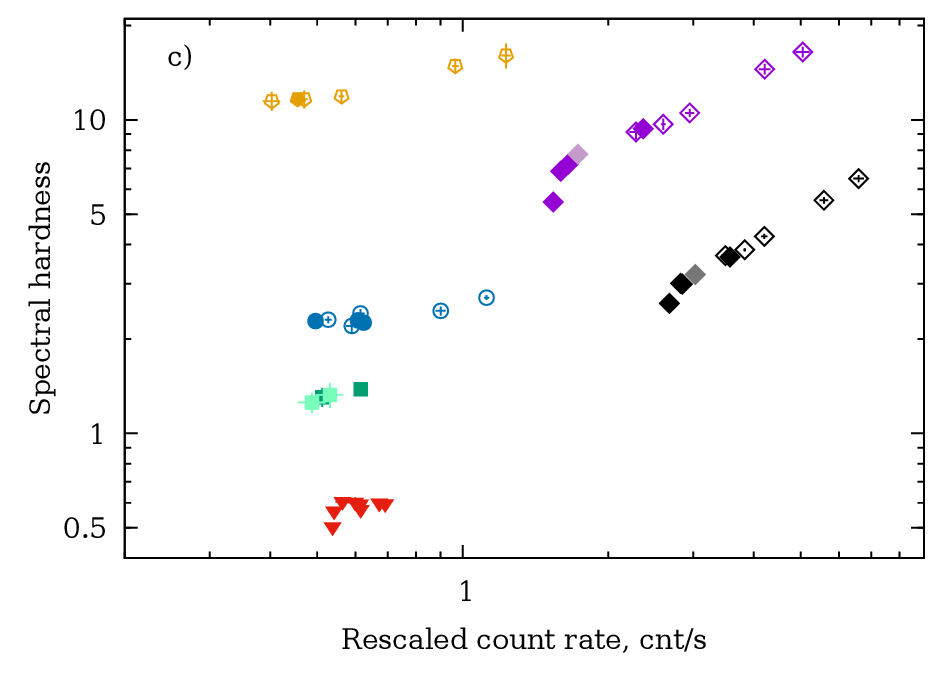}
\caption{Count rate in the 1--10~keV band (\textit{top panel}) and hardness ratio between the 1-10~keV and 0.3-1~keV bands (\textit{middle panel}) against the QPO frequency, and hardness against count rate (\textit{bottom panel}). To account the difference in distances to the ULX host galaxies, the count rates were rescaled to the distance of NGC\,5408. Observations of \6946 and \M82 where the significance of the QPO is less than $3\sigma$ (Table~\ref{tab:pds_fit}) are shown by lighter colours. Observations where we did not detect QPO are indicated by unfilled data points. Error bars (comparable to the point size) represent only intrinsic errors of the count rates, uncertainties of the distances are not included. Error bars of the QPO frequencies are omitted.}
\label{fig:qpo_flux}
\end{figure}

For all the observations we calculated the fractional rms variability \citep[$F_{rms}$,][]{Edelson2002,Vaughan2003} 
as a more accurate measure of the intrinsic variability of the source. Being an integral quantity, $F_{rms}$ does not depend on the specific shape of the PDS and expresses the total source variability within a certain frequency range. It is insensitive to the model uncertainties and can be obtained also for those PDS where we were unable to constrain the shape of the continuum. The fractional variability is defined as the root mean square of the light curve corrected by the effects of random-noise measurements and normalized to the mean source count rate.  

To ensure uniformity we obtained $F_{rms}$ from the 1000~sec length segments of combined pn+MOS light curves (this corresponds to the frequency range of $10^{-3} - 0.2$~Hz) for every observation regardless the QPO frequency. Such relatively long segments were chosen in order to include, at least in part, variability at frequencies below the break. Since the number of segments goes from 6 (in the worst case) to $\sim 100$, we were able to compute the sample mean and its standard deviation. Results are presented in Table~\ref{tab:frms}.

In Fig.~\ref{fig:qpo_rms} we present the fractional variability as a function of the QPO centroid frequency. The $F_{rms}$ (including contributions from both the continuum and QPO) is observed to decrease as the QPO frequency increases in all the cases. A power law fitting these data $F_{rms}\propto {f_q}^{-\gamma}$ yields $\gamma=0.31\pm0.03$ (1-$\sigma$ error) for \5408 and \6946. For \1313 we obtained a similar index $\gamma=0.34\pm0.03$ but the variability level shown by this source is significantly lower.

Although in the fractional variability both QPO and FTN contribute, we note that the observed negative correlation is provided mainly by the FTN, whose level monotonically and rapidly drops as the QPO frequency increases (Table~\ref{tab:pds_fit}). At the same time, variations of the power due to the QPO peak appear to be random (parameter $R_q$ in Table~\ref{tab:pds_fit}).

\M82 stands out from the other sources. It shows a shallower slope $\gamma=0.17\pm0.05$ for 32'' aperture (and $0.18\pm0.12$ for 12'') and a lower variability level. We suppose that such a difference may be due to the contamination of the source flux by diffuse emission from the host galaxy and neighbouring sources. This emission being less variable may `dilute’ the source variability. This idea is confirmed by the difference between the $F_{rms}$ values obtained from 32'' and 12'' radius apertures. In the case of a single isolated source, the fractional variability would not depend on the aperture 
size, since the absolute rms is in the numerator and the mean count rate is in the denominator change proportionally. But in the figure one can see that the smaller aperture gives higher $F_{rms}$ values because it collects less of the contaminating emission.

By taking a smaller aperture in \M82, one can reduce only the contamination by diffuse emission from the host galaxy. However much of the contaminating flux comes from the ULX pulsar M82~X-2. Sources X-1 and X-2 are only marginally resolvable by MOS detectors and completely unresolvable by pn. \cite{Pasham2013M82X1QPO} carried out surface brightness modelling of the MOS images from the same five \XMM\ observations we present here, and compared the luminosities of both ULXs. They found that X-2 was twice brighter than X-1 during the observation ObsID 0112290201 (the lowest data point in  Fig.~\ref{fig:qpo_rms}, that we excluded from the fit) and was fainter by a factor of 1.5--2 during the rest of observations. Since both sources are highly variable, \cite{FengKaaret2007M82}  and \cite{Pasham2013M82X1QPO} analysed the PDS from different areas near X-1 and X-2 in order to distinguish which source is the origin of the QPO. It has been confirmed that the QPO peak observed above 30~mHz is produced by M82~X-1. 
Source X-2 also showed  a QPO-like feature near 3~mHz but, above 10~mHz, its power spectrum is dominated by Poisson noise \citep{Feng2010M82X2QPOdiscov}. Thus, taking into account the contamination from both the diffuse emission of the galaxy and M82~X-2, we conclude that the true location of \M82 in Fig.~\ref{fig:qpo_rms} should be between \5408 and \1313, i.e. 2--3 times higher than it already is.

\begin{figure*}
\includegraphics[angle=0,width=0.45\textwidth]{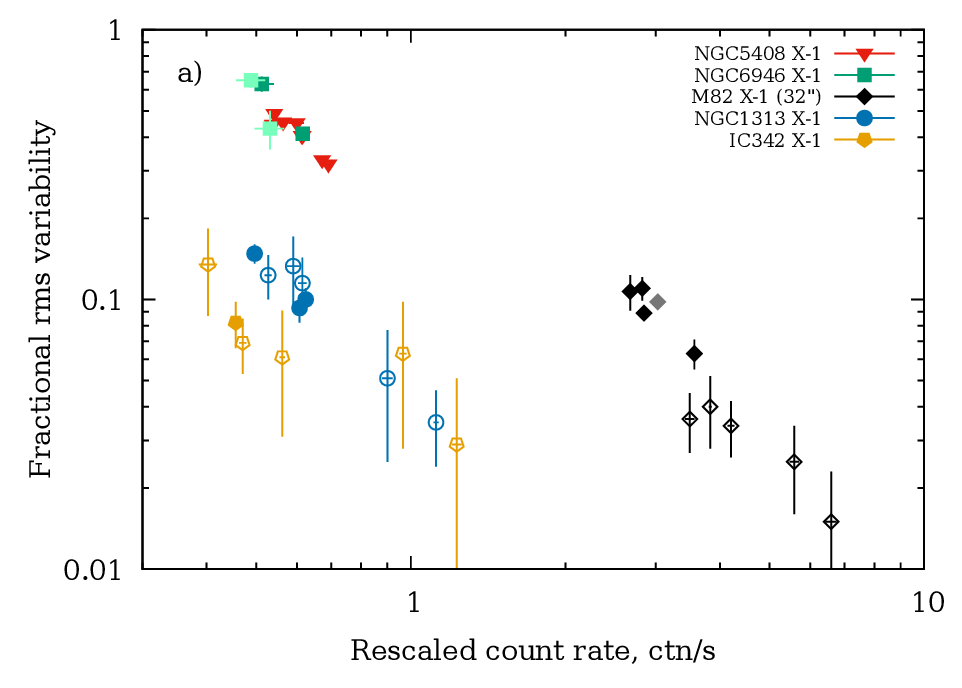}	
\includegraphics[angle=0,width=0.45\textwidth]{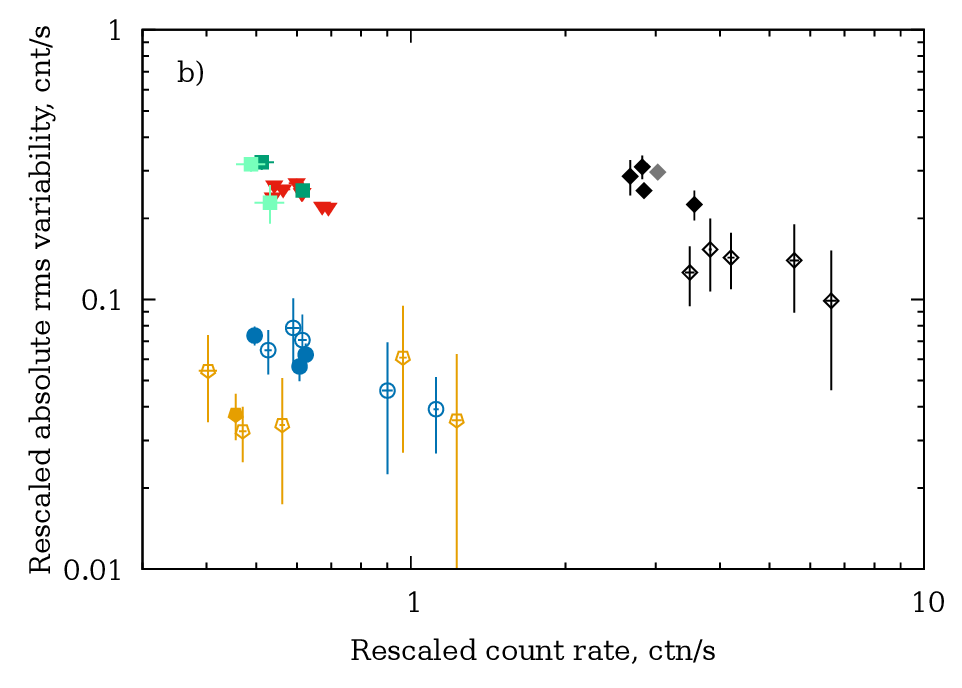}	
\caption{Fractional rms variability (\textit{left panel}) and absolute rms variability (\textit{right panel}) as a function of the count rate. The absolute rms variability and count rates are rescaled to the distance of \5408. Observations of \6946 and \M82 where the significance of the QPO is less than $3\sigma$ (Table~\ref{tab:pds_fit}) are shown by lighter colours. Observations where we did not detect QPO are indicated by unfilled data points.}
\label{fig:rms_flux}
\end{figure*}

In Fig.~\ref{fig:qpo_flux}a and Fig.~\ref{fig:qpo_flux}b we show the source count rate and spectral hardness as a function of the QPO centroid frequency. To be able to compare the luminosities of the ULXs located in different galaxies, we rescaled the count rates to the distance of \5408 (the distances to the host galaxies are provided in Table~\ref{tab:data}). Nevertheless, it should be noted that the (corrected) count rate (hereafter called apparent luminosity) represents only a fraction of the real (intrinsic) luminosity because it depends on a particular shape of the source spectrum and also does not account for possible beaming and absorption column which differs between sources and also can change over time \citep{Middleton2015atomic}. An increasing trend of the source apparent luminosity with the QPO frequency is seen in the figure. However this trend is relatively weak: if the frequency increases by ten times, the count rate changes by 40\% only. We obtained power-law slopes of $0.16\pm0.04$ for \5408, $0.12\pm0.04$ for \6946, $0.01\pm0.07$ for \M82 and $0.17\pm0.03$ \1313 (1-$\sigma$ errors presented).

The ULXs analysed in this paper are very different in spectral hardness (Fig.~\ref{fig:qpo_flux}b and \ref{fig:qpo_flux}c). It can be seen in Fig.~\ref{fig:qpo_flux}b that the hardness ratios of different sources are correlated with the frequency range in which the QPO appears in a particular source: \IC342 has the hardest spectrum among the ULXs of our sample and shows the highest QPO frequency (however this source has considerable absorption column, \citealt{Gladstone2009,Pintore2017}, which may affect the observed hardness ratio). The softest ULXs, \6946 and \5408, show the lowest frequencies, whist \1313 having an intermediate hardness shows a QPO at intermediate frequencies. At the same time, the variations of the QPO frequency between different observations of the same ULX almost do not change its spectral hardness. Fitting the hardness-frequency relation using a powerlaw, we obtained indices of $0.02\pm0.05$ for \5408, $0.047\pm0.015$ for \6946, $0.01\pm0.15$ and $-0.02\pm0.24$ for 32'' and 12'' apertures of \M82 respectively and $-0.002\pm0.011$ for \1313 (1-$\sigma$ errors). Only the slope of \6946 is different than zero by more than one sigma. This source shows a weak growth of the hardness with the QPO frequency (Fig.~\ref{fig:qpo_flux}b). In the case of \1313 and \IC342 the hardness does not change with the QPO frequency, however, it appears to be higher in the brightest observations of these sources where the QPO have not been detected (Fig.~\ref{fig:qpo_flux}c). 

In transient BHBs the spectral hardness is known to significantly change during an outburst and anti-correlates with the QPO frequency. In other words, a source becomes softer as the frequency increases \citep{Vignarca2003,ShaposhnikovTitarchuk2009}. The lack of this correlation in ULXs is one more argument supporting the idea of the difference between sub-Eddington BHBs and ULXs (see Sec.~\ref{discussion}). 

In the case of \M82 the contaminating flux makes it difficult to determine its true luminosity and hardness. However, despite the contamination, \M82 is an extremely bright ULX reaching an X-ray luminosity of $1.6\times10^{41}$ erg s$^{-1}$ \citep[from {\it Chandra} data]{Chiang2011M82}. \cite{FengKaaret2007M82} estimated unabsorbed luminosities of $1.3\times10^{40}$ and $1.7\times10^{40}$ erg s$^{-1}$ for data analysed from ObsIDs 011229020 and 0206080101, respectively. \5408 had only $7.8\times 10^{39}$ egs s$^{-1}$ during its brightest data set from ObsID 0653380301 \citep{Sutton2013}. Thus, even being relatively faint in the observations presented in Fig.~\ref{fig:qpo_flux}a, \M82 remains at least twice brighter than \5408. However it can not be discarded that the increase of the hardness with the count rate seen in Fig.~\ref{fig:qpo_flux}c might be caused by a brightening of the pulsar M\,82~X-2, which has a hard spectrum \citep{Brightman2016M82X2}. 
 
We have considered above only observations with QPO. Although the lack of QPOs in the rest of data sets in some cases might be caused by insufficient exposure time, we did not detect any specific features like flat-topped or power-law noise in their power spectra. The observations without QPO show some power above the level of Poisson noise, however, the shape of their PDS is rather random. For these observations we also calculated the fractional variability (Table~\ref{tab:frms}). 

In Fig.~\ref{fig:rms_flux}a we plot the fractional variability (including contribution from both continuum and QPO) versus the rescaled count rate for all analysed observations regardless the presence of the QPO. The $F_{rms}$ is seen to decrease with the apparent luminosity (Fig.~\ref{fig:qpo_rms} and \ref{fig:qpo_flux}a). The absolute rms variability (accounting for the measured noise) also decreases (Fig.~\ref{fig:rms_flux}b). The simultaneous reduction of both quantities indicates that the sources become intrinsically less variable as the luminosities increase. Also Fig.~\ref{fig:rms_flux}b suggests that the linear and positive relation between the absolute rms variability and flux obtained for time scales of a few hours \citep{Heil2010,Hernandez-Garcia2015} may be violated on longer time scales (months-years).

It is worth to mention that observations with and without QPO form a continuous sequence in Fig.~\ref{fig:rms_flux} in which the `lacking-QPO’ observations occupy the brightest 
part of the tail. This may suggest that each source has some luminosity threshold, above which the mechanism producing QPO and FTN breaks down and the variability almost disappears.

\section{Discussion} \label{discussion}
We have carried out a timing analysis of five ULXs that are known to show QPO and have found that their properties are very similar. In each source, the QPO frequency correlates negatively with the fractional rms variability and positively with the apparent luminosity. 
In general, sources with softer spectra show QPO at lower frequencies, however, a clear correlation between the spectral hardness and QPO frequency has been detected only in the case of \6946. We also confirmed that the QPO in the PDS always appears together with the FTN.

Sub-Eddington black hole binaries show QPO accompanied by FTN (the so-called type-C QPO) being in their hard intermediate state \citep{Belloni2016review}. Typical frequencies of the type-C QPO observed in stellar mass black holes are between 0.1 and 10~Hz (e.g. GX\,339-4 --- \citealt{Motta2011GX339QPO}; XTE\,J1550-564 --- \citealt{Rao2010XTEJ1550QPO}). It has been proposed that lower QPO frequencies in ULXs may be due to more massive black holes in these systems \citep{Strohmayer2009NGC5408IMBH}: if the frequency scales inversely proportional to $M_{BH}$, ULXs should contain black holes with masses of 1000--10000~$M_\odot$ \citep{Strohmayer2009NGC5408IMBH,Rao2010NGC6946QPOdiscov,DeMarco2013}. However, the identification of the QPO seen in ULXs with type-C implies that all observed spectral and variability properties must be consistent with the hard intermediate state of BHBs. But this is not the case. It is well-known that the energy spectra of ULXs are not alike the typical for BHBs \citep{Gladstone2009}. The spectra of ULXs when fitted with Comptonization models, yield a thicker and colder Comptonizing medium. The spectral hardness of BHBs is anti-correlated with the QPO frequency \citep{ShaposhnikovTitarchuk2009}. But in ULXs we have found that the hardness increases with the QPO frequency only in the case of \6946. The shape of the PDS also differs. In ULXs the QPO peak is located closer to the break \citep{Middleton2011}; we found mean values of $f_q/f_b\sim 3$ in ULXs versus $f_q/f_b\sim 10$ in BHBs \citep{Wijnands1999QPObreak}. Finally, ULXs show a larger spread of QPO frequencies (with \6946 having the lowest $f_q\approx 9.4$~mHz versus \IC342 having the highest $f_q\approx 654$~mHz) which, in the case of the mass scaling argument, would suggest a significant difference in the black hole masses from the ULXs. However, the ULXs analyzed in this paper show a similar behaviour and, probably, have a common underlying super-Eddington accretion nature. Indeed, the discovery of strong relativistic outflows in \5408 and \1313 \citep{Pinto2016Nat} argues in favour of super-Eddington accretion in these sources. 

It has been recently shown that some ULXs are pulsating neutron stars. At the moment five ULX pulsars have been found \citep{Bachetti2014Nat,Furst2016pulsP13,Israel2017pulsP13,Israel2017pulsNGC5907,Kosec2018NGC300ULX1}. Nevertheless, the analysis of a wide sample of ULXs (including \5408, \M82 and \1313) did not detect pulsations \citep{Doroshenko2015puls}. An explanation for the lack of pulsations has been proposed \citep{Walton2018puls}, however, we consider it unlikely that all known ULXs eventually will turn out to be neutron stars \citep{King2001ULX,King2017PULX,MiddletonKing2017ULXbeaming}. All the ULX pulsars have harder X-ray spectra than what is average in ULXs \citep{Pintore2017}. Many models have been suggested for the ULX-pulsars including magnetic fields of different strengths \citep{BaskoSunyaev1976limitlum, Mushtukov2015}. One of them requires geometrical beaming in order to violate the Eddington limit \citep{Kawashima2016ULXpulsmodel}.

The structure of supercritical accretion discs was first described by \cite{ShakSun1973}. If the matter supply in the outer boundary of the disc significantly exceeds the Eddington rate $\dot{m}_0=\dot{M}_0/\dot{M}_{Edd}\gg 1$ where $\dot{M}_{Edd}=48\pi G M/c\kappa$ (assuming a radiative efficiency $\eta=1/12$) and $\kappa$ is Thomson opacity, then there is a radius where the force of the radiation pressure becomes equal to the gravity force. This radius is called \textit{spherization radius} and it limits the supercritical area of the disc. Below the spherization radius the disc becomes geometrically thick with $H/R\sim 1$ and it is where the wind is launched. Since the wind possesses an angular momentum, it is supposed to form a funnel with some opening angle. The size of the supercritical area is proportional to the initial accretion rate $R_{sp}=(5/3)\dot{m}_0 R_{in}$ \citep{Poutanen2007}. The inner radius of the disc $R_{in}$ is usually assumed to be $3R_g$, where $R_g={2GM/c^2}$ is the Schwarzschild radius. Due to the outflow, below $R_{sp}$ the amount of matter accreting through the disc linearly decreases 
with radius $\dot{m}(R) = \dot{m}_0\cdot (R/R_{sp})$. This basic concept proposed by \cite{ShakSun1973} and developed in more detail in works by \cite{Lipunova1999} and \cite{Poutanen2007} have been well reproduced by MHD simulations \citep{Ohsuga2005,Ohsuga2011}.

Below we suggest an interpretation which connects the observed ULX variability properties and the PDS shape in the framework of super-Eddington accretion.

\subsection{Lyubarskii’s proposal}

The variability of X-ray sources is usually interpreted in terms of the mass accretion rate fluctuations arising from random fluctuations of viscosity at different accretion disc radii and propagating inwards. This mechanism was proposed by \cite{Lyubarskii1997} to explain power-law PDS of BHBs. It is supposed that the viscosity varies in all time and spatial scales. The variations of viscosity at one radius change the accretion rate at smaller radii. However, because of the diffusive nature of accretion, only variations on time scales longer than \textit{accretion time} (or \textit{viscosity time}) can propagate whilst the faster variations will be damped. The accretion time increases with radius as: 
\begin{equation}
t_a(R) = \left[\alpha \left(H\over R\right)^2 \Omega_K(R) \right]^{-1}
\label{eq:ta}
\end{equation}
where $\alpha$ is the viscosity parameter \citep{ShakSun1973}, $H$ is the half-thickness of the disc and $\Omega_K$ is the Keplerian frequency; so the outer regions of the disc produce the slowest variations. As the matter passes through the disc, more and more rapid variations are superimposed on these slow ones. Finally, when the matter reaches the inner disc radius, where most of the energy is released, it bears the variations accumulated from all possible disc radii which makes the shape of the resultant PDS power-law. The slope of the PDS is determined by the radial dependence of the amplitude of the viscosity fluctuations. In particular, if the amplitude is the same at all radii then the power spectrum should be $P\propto f^{-1}$.

\cite{Uttley2001rmsflux} have pointed out that additionally to the power-law PDS the mechanism proposed by Lyubarskii also predicts a linear relation between the absolute rms variability and flux, which cannot be explained by other models (for example shot noise model from \citealt{Terrell1972shotnoise}). As it latter became clear, the linear rms-flux relation is a very common feature of accreting systems and may be a more fundamental characteristic of the variability than the power spectrum  or spectral-timing properties \citep{Uttley2005rmsflux}. The linear rms-flux relation is observed in binary systems with black holes and neutron stars \citep{Uttley2001rmsflux,Gleissner2004CygX1rmsflux,Uttley2004pulsarrmsflux}, active galactic nuclei \citep{Vaughan2003AGNrmsflux,Vaughan2005AGNNGC4395} and cataclysmic variables \citep{Scaringi2012WDrmsflux}. This fact made the Lyubarskii’s model and its extensions \citep{King2004variability,Arevalo2006,Titarchuk2007variability,Ingram2011} very convincing. The discovery of the same behaviour in ULXs \citep{Heil2010,Hernandez-Garcia2015} suggests that this mechanism must work in these systems as well. Moreover, the Lyubarskii’s model was successfully applied to the variability of the supercritical accretion disc of \SS433 \citep{Revn2006,AtapinSS433var2015,Atapin2016pazh}.

\subsection{Flat-toped noise, QPO and spherization radius}

Pure power-law PDS predicted by \cite{Lyubarskii1997} for a standard accretion disc can be formed if the accretion time and the fluctuations amplitude smoothly change with radius. In the case of the supercritical disc this is probably not true. Above the spherization radius, where the energy release is not so strong, the structure of the supercritical disc is supposed to be similar to the standard geometrically thin disc with $H/R\sim 0.03-0.1$ \citep{ShakSun1973}. But below $R_{sp}$ the disc becomes thick $H/R\sim 1$ (a more precise model including advection, i.e. \citealt{Lipunova1999}, predicts a disc thickness of 0.6-0.7), which drastically reduces the accretion time. The matter passes a distance from $R_{sp}$ to the innermost orbit almost in the free-fall time. In such a case the spherization radius becomes a trigger, which regulates the matter supply into the supercritical area of the disc. We suppose that if the viscosity at $R_{sp}$ varies in a random and independent manner (``white noise'') as it was assumed for all the radii \citep{Lyubarskii1997}, this effect should produce a flat portion in the PDS  \citep{AtapinSS433var2015,Middleton2015model}.

\begin{figure}
\includegraphics[angle=0,width=0.45\textwidth]{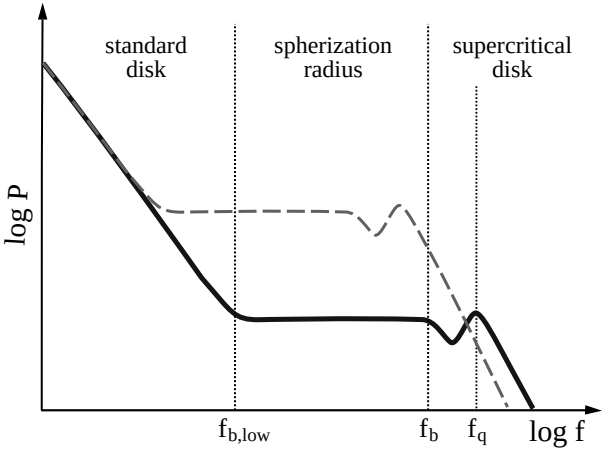}	
\caption{Sketch of the broadband power spectrum of the supercritical accretion disc. Frequencies $f_q$ and $f_b$ mark positions of the QPO and the break observed in ULXs analysed in this work (Fig.~\ref{fig:pds}). The $f_{b,low}$ denotes position of hypothetical low-frequency break lying outside the observable \citep{Atapin2016pazh,Atapin2017ASPC_SAO50conf} bandpass. We propose that all three characteristic frequencies are related to the size of the supercritical area of the disc ($R_{sp}$) which is determined by the accretion rate. The flat-toped noise (FTN) in the PDS occurs due to fluctuations directly at $R_{sp}$, while the power-law portions $f<f_{b,low}$ and $f>f_b$ originate from the $R\gg R_{sp}$ and $R<R_{sp}$ regions of the disc respectively. As the accretion rate increases, the characteristic frequencies shift toward lower values and the level of FTN becomes higher (power spectrum for the case of higher accretion rate is shown by dashed line).}
\label{fig:scheme}
\end{figure}

In the suggested model the FTN arises from fluctuations occurring at the spherization radius, and the break frequency $f_b$ in the PDS is determined by $t_a$ at $R_{sp}$. At higher frequencies the power spectrum is related to the fluctuations occurring in the supercritical part of the disc ($R<R_{sp}$). At these frequencies the PDS shape should be power-law as from \citet{Lyubarskii1997}, since the accretion time strongly decrease from $t_a(R_{sp})$ to $t_a(R_{in})$. These power-law portions with index $\beta$ (Table~\ref{tab:pds_fit}) are clearly seen in Fig.~\ref{fig:pds}. Frequencies much lower than the break and FTN should correspond to the subcritical part of the disc ($R\gg R_{sp}$), here the PDS can become power-law again with another (low-frequency) break (Fig.~\ref{fig:scheme}). \citet{Atapin2016pazh} discovered such a low-frequency break in \SS433 and also found some hints of the power law below $10^{-4}$~Hz in the ULXs \5408 and \6946 \citep{Atapin2017ASPC_SAO50conf}.

Writting $R_{sp}$ as a function of the initial accretion rate, the break frequency can be estimated as (Hz):
\begin{equation}
f_b\sim t_a^{-1}(R_{sp}) \approx 6.0\times 10^{-3} \alpha_{0.1}\dot{m}_{300}^{-3/2}m_{10}^{-1}  
\label{eq:fb}
\end{equation}
where the disc thickness below $R_{sp}$ is assumed to be $H/R=0.7$ \citep{Lipunova1999}, $\alpha_{0.1}=\alpha/0.1$, the mass accretion rate $\dot{m}_{300}$ is in units of $300\dot{M}_{Edd}$ and the black hole mass is in units of $10M_\odot$.
Using $\dot{M}=300\dot{M}_{Edd}$, $M_{BH}=10M_\odot$ and $\alpha=0.1$, one obtains the break at 6\,mHz, the same as observed in \5408 and \6946 (Table~\ref{tab:pds_fit}). The accretion rate of $300\dot{M}_{Edd}$ seems reasonable for ULXs because the same value is  typical for \SS433 (assuming $M_{BH}\approx 10M_{\sun}$). Other ULXs showing breaks at higher frequencies may have smaller accretion rates or black hole masses.

The QPO in this model may arise due to instabilities of the accretion flow in the supercritical area of the disc. Below the spherization radius the gravity is dominated by the radiation pressure. The matter is supported by the radiation field; otherwise it could approach the black hole or its surroundings in the free-fall time scale. We suppose that in such a case the QPO frequency may be related to the Keplerian time at the spherization radius:
\begin{equation}
f_q \approx P_K^{-1}(R_{sp}) ={\Omega_K(R_{sp})\over2\pi} \approx 2.0\times 10^{-2} \dot{m}_{300}^{-3/2}m_{10}^{-1}  
\label{eq:fq}
\end{equation}
This formula yields a constant ratio $f_q/f_b = (2\pi (H/R)^2 \alpha)^{-1} \sim 3$ (for $H/R=0.7$ and $\alpha=0.1$) which is in good agreement with observations (Sec. 3.1). For typical $\dot{M}_0=300\dot{M}_{Edd}$ and $M=10M_\odot$ one obtains $f_q\approx 20$~mHz. QPOs at similar frequencies arising due to instabilities in the supercritical disc have been reproduced in radiation hydrodynamical calculations by \cite{Okuda2009}.

\subsection{Accretion rates and fractional rms variability}

As we have shown in Sec. 3.1, the QPO and break frequencies can vary by 3-5 times between different observations of the same source. We propose that such a behaviour is caused by variations of the initial accretion rate $\dot{m}_0$ (i.e. accretion rate supplied by the donor star to the outer disc edge). Other parameters like viscosity $\alpha$, inclination of the super-Eddington disc and opening angle of the wind-funnel (geometrical beaming) may also change and affect the characteristic time scales. To not to complicate the situation with $\alpha$ and all other parameters (which are not understood yet, but may vary a little) we discuss here only $\dot{m}_0$. We propose that the QPO frequency has to be anti-correlated with the accretion rate, i.e. as $\dot{m}_0$ increases, the spherization radius widens and the QPO and break frequencies decrease. Variations of the QPO frequency by 5 times (in each individual source) imply a variation of the accretion rate by $\approx 3$ times. Using eq.~(\ref{eq:fq}) and assuming the same black hole mass $M=10M_\odot$ for all the five ULXs (better estimates for the masses will be discussed below) we estimate $\dot{m}_0$ as $180-450$ for \5408, $180-520$ for \6946, $90-205$ for \M82, $50-120$ for \1313 and 30\,$\dot{M}_{Edd}$ for \IC342. If the real masses are lower then the accretion rates need to be higher. 

In Sec 3.2 we have shown that the sources become less variable as the QPO frequency increases. We found $F_{rms}\propto f_q^{-\gamma}$ with slopes $\gamma = 0.31-0.34$ for \5408, \6946, \1313 and $\gamma\approx 0.17$ for \M82 (Fig.~\ref{fig:qpo_rms}). In the case of \IC342 there is only one detection of QPO, and it is not possible to determine $\gamma$ for this ULX. For \M82 one cannot accurately measure the fractional variability because of the very crowded spatial environment. The real slope of \M82 could be steeper. Taking this into account we can obtain a relation between the fractional rms variability and the accretion rate: $F_{rms}\propto \dot{m}_0^{0.5}$ (for $\gamma\approx 0.3$).

Such a positive correlation between the fractional variability and the accretion rate is in good agreement with Lyubarskii’s proposal. Also a similar relation between the fractional variability and the break frequency was considered by \cite{Middleton2015atomic}. A major contribution to the $F_{rms}$ is due to the FTN for which its level is expected to depend on the accretion rate. In the model described above and illustrated in Fig.~\ref{fig:scheme} the level of FTN is determined by the low-frequency break $f_{b,low}$. As the accretion rate increases, the supercritical region of the disc increases and the time scales at $R_{sp}$ become longer: Thus the FTN together with QPO and both breaks move upward and to the left in Fig.~\ref{fig:scheme}. The total area under the PDS ($F_{rms}$) also increases. When the $\dot{m}_0$ decreases, the FTN goes to in the opposite direction and the $F_{rms}$ decreases.

In Fig.~\ref{fig:rms_flux}a we show the fractional variability of the observations where we did not detect QPO. In this figure the observations with and without QPO form a single sequence in which the `QPO-less’ observations show lower $F_{rms}$. This fact may imply that the accretion rate in the `QPO-less’ observations is significantly lower. We suggest that only \5408 and \6946 from the list of the five ULXs considered here have higher $\dot{m}_0$ because in these sources we always see a QPO (if one includes two observations of \6946 where the QPO is only marginally detected). In other ULXs (\1313, \M82\ and \IC342) the QPO disappears at higher frequencies when $\dot{m}_0$ and $F_{rms}$ decrease. Using observations with the lowest $F_{rms}$ (Table~\ref{tab:frms}) and the relation  
$F_{rms}\propto \dot{m}_0^{0.5}$ we can estimate the lowest accretion rate for \1313, \M82 and \IC342 as 7.1, 2.6 and 
$3.8\,\dot{M}_{Edd}$ respectively. However, these values might be underestimated. This result depends on the black hole mass. Nevertheless, in these three ULXs the QPO may disappear when accretion rate falls below some critical level.

In the following we explain the possible reasons for the presence of QPOs and FTN in the power spectra of only a few (five) ULXs but not in others. In Table~\ref{tab:pds_fit} 
we see that the QPO peak becomes broader as the frequency increases (Fig.~\ref{fig:fq_fwhm}). It is common for Lorentzian components that with increasing frequency the FWHM must increase as well \citep{Belloni2002}. We suggest that QPOs might disappear because of the broadening. Therefore, if $\dot{m}_0$ reduces then the QPO peak moves to higher frequencies, it broadens and disappears. Having a low enough accretion rate may be the reason of not observing FTN and QPO in some well-known ULXs. Through the study of the PDS of 16 sources, a sample of ULXs with low variability has been revealed by \cite{Heil2009}. More observations are needed to confirm that other ULXs really do not have QPOs, since the five ULXs with QPO studied here have been observed during long exposures.

\subsection{X-ray luminosity, spectral hardness and suggested black hole masses}

We found that the observed X-ray luminosity slightly increases with the QPO frequency (Fig.~\ref{fig:qpo_flux}a) as $L_X\propto f_q^{0.16\pm0.04}$ (for \5408, Sec. 3.2). Taking into account the relation~(\ref{eq:fq}) of $f_q$ and $\dot{m}_0$, it would result into a negative correlation between the accretion rate and luminosity. On the other hand, the bolometric luminosity of the super-Eddington accretion disc should increase with higher accretion rate (\citealt{ShakSun1973,Poutanen2007}; assuming the 
beaming factor equal to unity):
\begin{equation}
L_X=L_{Edd}(\ln{\dot{m}_0}+1)
\label{eq:Lx}
\end{equation}
We suppose that there is no conflict between observations and theory here. Part of the X-ray radiation originating from the innermost regions of the funnel must be scattered or absorbed in the wind. The stronger the outflow, the more the X-ray photons will be scattered. Since most of the matter supplied to the disc by the donor star is eventually ejected from the system \citep{ShakSun1973,Poutanen2007}, then the mass loss rate in the wind $\dot{m}_w$ has to be strongly dependent on $\dot{m}_0$. When $\dot{m}_0$ increases, the logarithmic factor $\ln{\dot{m}_0}$ has to increase, but the X-ray luminosity cannot overcome the scattering in the wind. Of course, it could depend on other parameters (for example, beaming), however, this may complicate the situation.

\begin{figure}
\includegraphics[angle=0,width=0.45\textwidth]{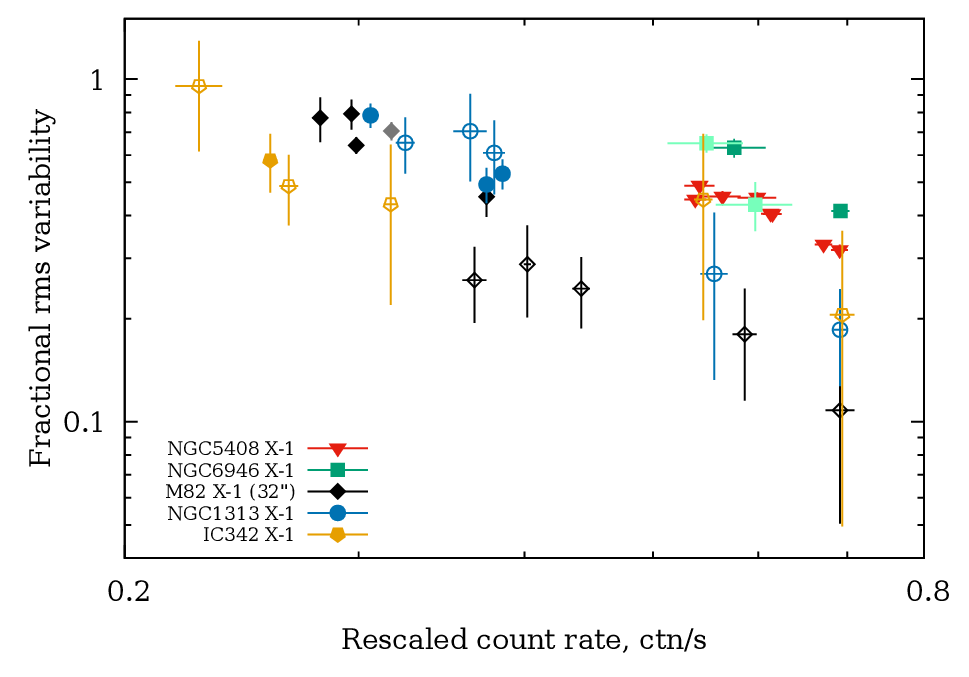}	
\caption{Similar to Fig.~\ref{fig:rms_flux}a but accounted for the difference in the accretion rate and black hole mass between the sources. The distances, masses and accretion rates are rescaled to those that in \5408. The legend is the same as in Fig.~\ref{fig:qpo_rms}. Observations without QPO are shown by unfilled data points. Lighter colours denotes observations of \6946 and \M82 with the QPO significance less than $3\sigma$.}
\label{fig:rms_flux_mass}
\end{figure}

Using the observed luminosities we can estimate the black hole masses because the X-ray luminosity in eq.~(\ref{eq:Lx}) is proportional to $L_{Edd}$ which scales with the mass. The result might be affected by $\dot{m}_0$ through both the logarithmic factor and the scattering of X-ray photons in the wind. Therefore, we used only the brightest observation in Table~\ref{tab:data} and Fig.~\ref{fig:rms_flux} for each ULX  to reduce uncertainties due to scattering. Expressing the masses in units of the mass from the BH in \5408, one obtains 0.9, 9.5, 1.6 and 1.8 for \6946, \M82, \1313 and \IC342 respectively. So, all the ULXs except \M82 have similar masses. It is clear that more accurate assessments should also consider other parameters, namely beaming and inclination. Moreover in the case of \5408 and \6946 luminosities and masses might be underestimated because in these sources we see QPO even in the brightest observations and the scattering may be significant.

The spectral hardness also can depend on accretion rate. In Fig.~\ref{fig:qpo_flux}b we showed that the sources show very different hardness, and that it correlates with the frequency of the QPO appears in each source. The hardest one is \IC342 \citep{Shidatsu2017IC342} having the highest QPO frequency and, probably, the lowest accretion rate. The softest ones, \5408 and \6946 show the lowest QPO frequencies and seem to have the highest accretion rate. This could imply that the hardness increases as the accretion rate decreases. This effect might be due to, for example, the optically thick wind which having a relatively soft spectrum \citep{Poutanen2007} may cover the innermost hottest regions of the accretion disk emitting hard X-rays \citep{Middleton2011,Sutton2013}. Nevertheless, it is still unclear why the hardness does not change with the QPO frequency.  We found small changes only in \6946 (Fig.~\ref{fig:qpo_flux}b). However in the brightest observations of \1313, \IC342 and \M82 (without QPO, Fig.~\ref{fig:qpo_flux}c) the hardness appears to be higher. 
Also we do not understand why \5408 and \6946 differ in hardness, despite they have similar accretion rate and masses. It might be related to different black hole  spin or orientation of the systems but we do not consider these parameters for simplicity. The effect of inclination on spectral hardness is discussed in \cite{Middleton2015model}.

In Fig.~\ref{fig:rms_flux_mass} we show the fractional rms variability versus the rescaled count rate similar to Fig.~\ref{fig:rms_flux}a but corrected for the difference in the black hole mass and accretion rate. Both quantities are rescaled to \5408. To recalculate the luminosity we used the masses obtained above and eq.~(\ref{eq:Lx}). For the fractional variability we used the relation $F_{rms}\propto f_q^{-\gamma}$ with $\gamma\approx 0.31$ for \6946, $\gamma\approx 0.17$ for \M82 and $\gamma\approx 0.34$ for \1313 and \IC342. In Fig.~\ref{fig:rms_flux_mass} the ULXs become much closer than they are in Fig.~\ref{fig:rms_flux}a, especially \M82, \1313 and \IC342. However, some differences between the sources in this figure still remain. It is possible that other parameters, that we did not take into account, are important.

Obviously, this interpretation is just a suggestion; the masses from all ULXs can depend on other parameters like beaming, orientation, etc. Furthermore, the wind itself may produce the variability \citep{Middleton2011,Middleton2015model} which makes the picture much more complicated. We only follow what \cite{ShakSun1973} and \cite{Lyubarskii1997} found in order to explain the behaviour of the observed variability from ULXs. Those studies allowed us to connect the flat-topped noise, QPO, $F_{rms}$, luminosity and spectral hardness with the spherization radius and accretion rate.

\section{Conclusions}
In this paper we have analyzed the power spectra of five ULXs that are known to show considerable amount of fast variability and QPO. We found that, despite the fact that ULXs significantly differ in the spectral hardness and QPO frequencies, their variability properties are quite similar. In all the cases the QPO is accompanied by flat-toped noise whose level is anti-correlated with the QPO frequency. As the frequency increases, the sources become brighter and less variable. 

We propose that this behaviour can be explained in terms of supercritical accretion from \cite{ShakSun1973} and propagating fluctuations from \cite{Lyubarskii1997}. Most of the observed variability properties are governed by the accretion rate which regulates the size of the supercritical area of the disc (spherization radius). Both QPO and FTN are related to the accretion time at $R_{sp}$. As the accretion rate increases, the spherization radius increases and the FTN and QPO move toward lower frequencies and upwards. The level of FTN as well as the fractional variability increase. The outflow from the disc also increases, which makes the apparent X-ray luminosity lower due to scattering in the wind. In opposite situation, when we do not observe the QPO and FTN, the $\dot{m}_0$ might be too small (but anyway super-Eddington). This could be the reason why one cannot observe QPO in other ULXs.

We infer that \5408, \6946, \1313 and \IC342 may contain black holes of similar masses, whilst the black holes in \M82 may be $\approx10$ times more massive. From the QPO frequencies we conclude that the highest accretion rates are found in \5408 and \6946 ($\sim 200-500\,\dot{M}_{Edd}$) and the lowest in \IC342 ($\sim30\,\dot{M}_{Edd})$. In the case of observations where QPO have not been detected, the accretion rate may be even lower.
		
\section*{Acknowledgements}
We thank the anonymous referee for careful reading of the paper and useful remark. This work is based on the data from the \XMM, an ESA Science Mission with instruments and contributions directly funded by ESA member states and the USA (NASA). The research was supported by the Russian RFBR grants 18-32-20214, 19-02-00432. MCG acknowledges support provided by the European Seventh Frame-work Programme (FP7/2007-2013) under grant agreement n$^{\circ}$ 312789 and GA CR grant 18-00533S and also  funding from ESA through a partnership with IAA-CSIC (Spain).


\end{document}